\documentclass[
reprint,aps,prb]{revtex4-2}

\usepackage{hyperref}

\usepackage{xcolor}
\usepackage{graphicx}
\usepackage{bm}
\usepackage{mathtools}
\usepackage{amsfonts}
\usepackage{physics}
\usepackage{cleveref}

\usepackage{ragged2e}
\usepackage[caption=false]{subfig}
\usepackage{tabularx}
\usepackage{tikz}

\DeclarePairedDelimiter\ceil{\lceil}{\rceil}
\DeclarePairedDelimiter\floor{\lfloor}{\rfloor}

\usepackage{orcidlink}
\allowdisplaybreaks

\begin{document}

\title{Algebraic structure of finite-size Bethe solutions in the Heisenberg chain}

\author{O. R. Bellwood \,\orcidlink{0000-0002-0691-4175}}
\email{oliver.bellwood@oist.jp}
\affiliation{Quantum Engineering and Design Unit, Okinawa Institute of Science and Technology Graduate University, Onna-son, Okinawa 904-0495, Japan}

\author{W. J. Munro \,\orcidlink{0000-0003-1835-2250}}
\affiliation{Quantum Engineering and Design Unit, Okinawa Institute of Science and Technology Graduate University, Onna-son, Okinawa 904-0495, Japan}

\date{\today}

\begin{abstract}
The spin-$1/2$ Heisenberg antiferromagnetic chain is the canonical example of an integrable quantum many-body model. Despite its exact solvability, explicit finite-size solutions are typically only accessible via numerical evaluation of the Bethe ansatz equations. Here, we analyse the algebraic structure of the exact, symbolic ground states for chains up to twelve sites using the Bethe ansatz. We find that the algebraic degree of the ground state magnon phases and Bethe roots scale with respect to system size at a different rate to the ground-state energy and wavefunction coefficients. The ground-state Bethe roots and energy are found to become Galois unsolvable for ten or more sites, thus demonstrating the lack of a symbolic analytical solution in a quantum integrable model due to algebraic complexity. 
\end{abstract}

\maketitle

\textit{Introduction}.---
Integrable quantum many-body systems occupy a central position in condensed matter and mathematical physics. Among the most celebrated examples is the one-dimensional, spin-$1/2$ antiferromagnetic Heisenberg model (Heisenberg chain), whose exact solution via the Bethe ansatz inaugurated the modern theory of quantum integrability \cite{bethe1931theorie, hulthen1938uber,BaxterBook,Takahashi_1999,doi:10.1142/S0217751X10049803}. Within Bethe's framework, the many-body eigenproblem reduces to a set of coupled nonlinear equations for the momenta and interaction terms between spin-wave quasiparticles; this methodology is termed the ``coordinate Bethe ansatz'' \cite{korepin1993quantuminversescatteringmethod,BA1,BA2,Staudacher}. These equations determine the finite-size spectrum exactly and, in principle, furnish complete information about the corresponding eigenstates. It was later shown that the Bethe ansatz admits an entirely algebraic treatment \cite{Yang_1966}, whereby one never has to consider the quasiparticle picture of the Heisenberg chain excitations; this is termed the ``algebraic Bethe ansatz'' \cite{FADDEEV_1995,faddeev1996algebraicbetheansatzworks,NEPOMECHIE_1999}, which has seen  extensive interest in the past half century.

In spite of the formal exactness of the Bethe ansatz, explicit finite-size solutions are rarely available in symbolic form. The Bethe ansatz equations are almost invariably solved numerically as they allow arbitrarily high precision solutions for moderate system sizes \cite{FCAlcaraz_1987,10.1063/1.3216474,DBiegel_2002}. However, the numerical solutions obscure a large amount of the mathematical structure of the physical eigenstates. Conceptually, quantum integrability does not guarantee explicit symbolic solvability, only the existence of exact defining equations \cite{Caux_2011,Larson_2013,Batchelor2015,Links_2021,ISAAC2026105837,Links_2026}. In the context of algebraic geometry, this is exemplified by the emergence of Galois unsolvable solutions (i.e. solutions unable to be expressed in radicals) \cite{Jiang_2018,Jacobsen2019Torus}. Solutions to the Heisenberg chain are analytically exact yet can lack an elementary closed form if they are Galois unsolvable. However, determination of exactly \textit{when} Galois unsolvability emerges in the Heisenberg chain is impossible to determine from numerical Bethe solutions alone. As quantum integrability and exact solvability are used synonymously, determining the symbolic solutions aids in sharpening these two concepts in the many-body literature, insofar as elementary algebraic solvability is not implied by quantum integrability.

More broadly, symbolic solutions of the Bethe ansatz equations allow the interrogation of fundamental aspects of the Heisenberg chain wavefunction \cite{Hagemans_2007,Hao_2013, Jiang_2018}. Presently, we attempt to address a rather innocuous question: what is the minimal algebraic structure of the Heisenberg chain ground state? For bipartite antiferromagnets, such as the finite Heisenberg chain with an even number of sites, the ground state is known to be a total spin singlet \cite{Lieb_Mattis_1962,auerbach}. Therefore, the wavefunction is representable in the basis of singlet dimer coverings, which is more compact than the computational basis \cite{TemperleyLieb1971,Saito_noncrossing,BEACH2006142,kirillov2009valencebond}. But is this the ``fundamental'' basis that governs the algebraic complexity of the ground state? 

Here, we derive the exact ground states up to twelve sites via the algebraic and coordinate Bethe ansatz. We find that the ground state quantities develop algebraic complexity with respect to system size at a different rate to the Bethe roots. We conjecture which integer sequences govern the algebraic degree of the Bethe roots and the ground state itself and comment on the physical significance of this algebraic structure to the representation of the wavefunction.
Furthermore, we find that the ground state quantities and Bethe roots become Galois unsolvable for chains of ten- or more sites, in agreement with similar findings from the six-vertex model \cite{Jacobsen2019Torus}.

\begin{figure}[t]
\centering
\hfill
\subfloat[\label{N2_fig}]{%
\includegraphics[width=0.45\columnwidth,trim={0 0.8cm 0 0}]{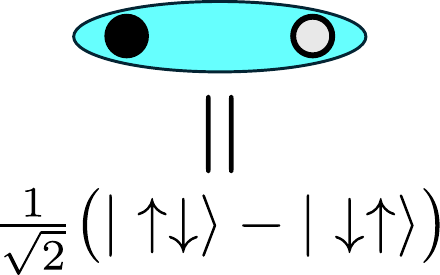}%
}\;\;\;\;
\subfloat[\label{N4_fig}]{%
\includegraphics[width=0.4\columnwidth,trim={0 0.8cm 0 0}]{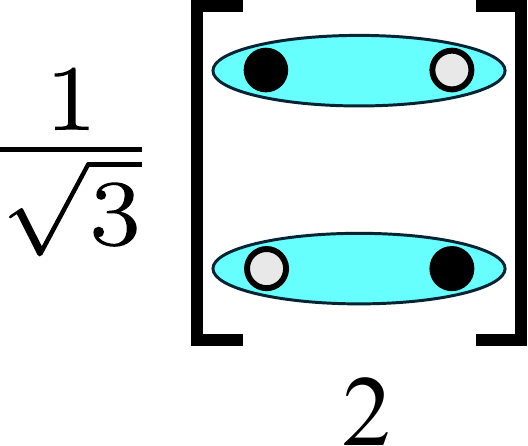}%
}\;\;\;\;
\subfloat[\label{N6_fig}]{%
\includegraphics[width=0.9\columnwidth,trim={0 0.8cm 0 0}]{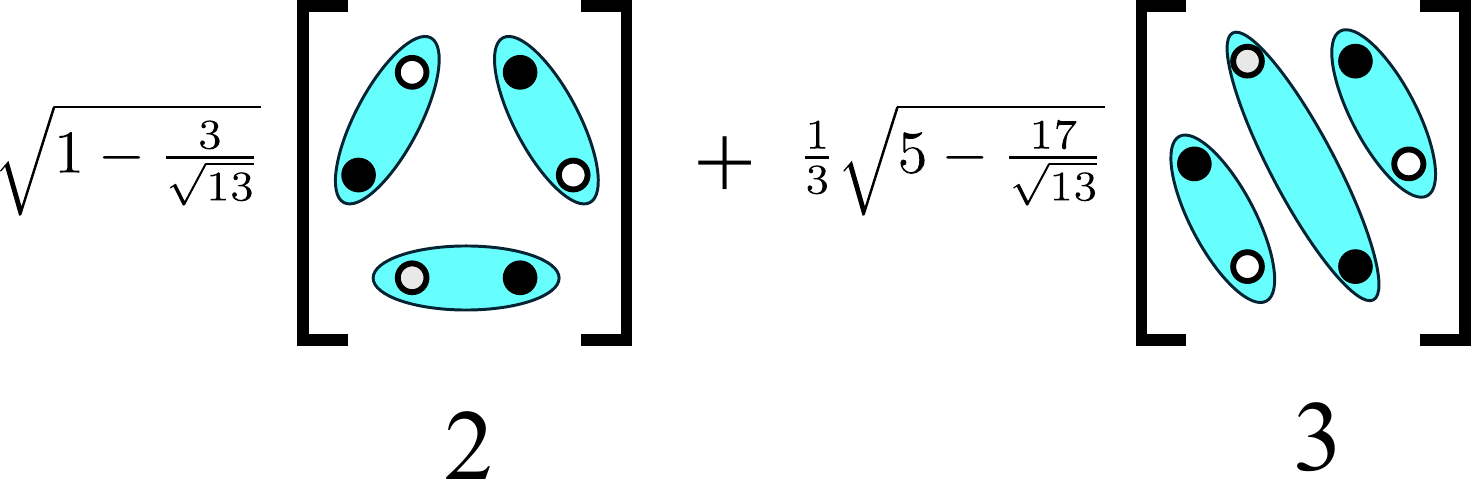}%
}\;\;\;\;
\subfloat[\label{N8_fig}]{%
  \includegraphics[width=0.9\columnwidth]{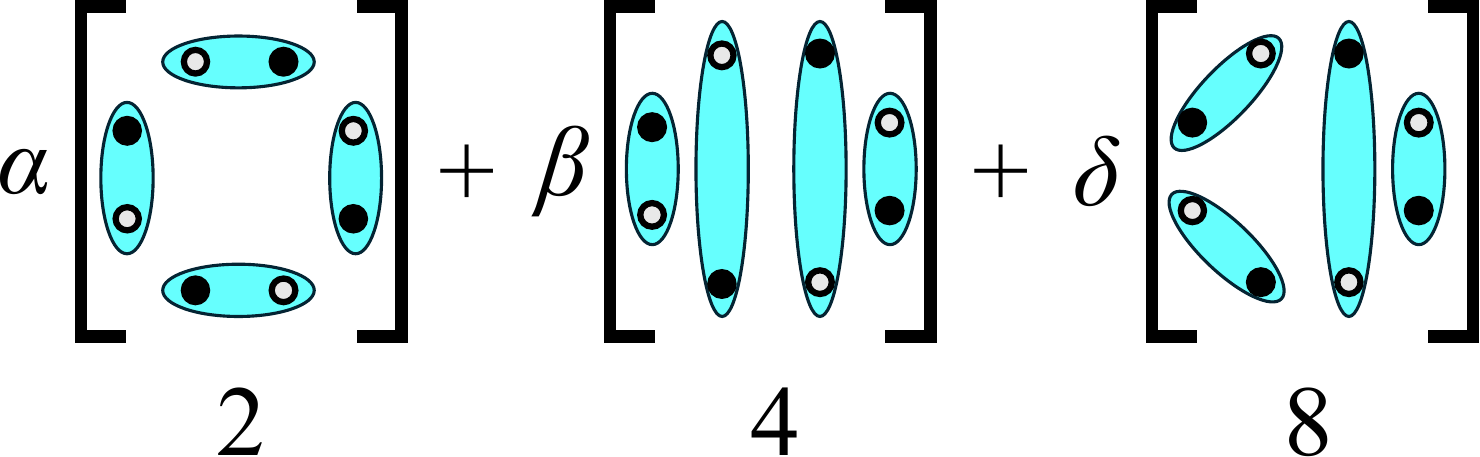}%
}
\caption{
 Schematic of the exact symbolic ground states of the two-, \mbox{four-,} six-, and eight-site Heisenberg chains in the non-crossing valence-bond basis. Square brackets denote a linear combination over valence-bond configurations reached via symmetry transforms, and the number below the square brackets is the multiplicity of such configurations. At eight sites (d), the (normalized) coefficients $\alpha\approx 0.30, \beta\approx 0.03,$ and $\delta\approx 0.11$ are algebraic numbers of degree six; notably, the unnormalized Bethe ansatz wave-function coefficients, \cref{coeffs}, are degree three despite the magnon cosine phases and momenta being degree six. The ten- and twelve-site ground states are given in the supplemental materials \cite{supp}. 
}
\label{Figs}
\end{figure}

\textit{Algebraic and Coordinate Bethe Ansatz}.---
We consider the finite-sized, spin-$1/2$ Heisenberg Hamiltonian
\begin{align} \label{Ham}
\hat{H} = J\sum_{i=1}^{N}\vec S_i \cdot \vec S_{i+1},
\end{align}
with $J=1$ and $\vec S_{N+1}=\vec S_1$. Restricting to even $N$, the antiferromagnetic ground state lies in the $S^z=0$ sector with $r=N/2$ down spins (magnons) \cite{Lieb_Mattis_1962}. A computational basis state with $r$ magnons exists in a subspace of dimension ${N}\choose{r}$, and is defined relative to the ferromagnetic reference state, $\ket{F} \equiv \ket{\uparrow...\uparrow}$, by the position of the down spins as $ \ket{n_1, ..., n_r} \equiv S^-_{n_1}...S^-_{n_r}\ket{F}$. An (unnormalized) eigenstate is written as $\ket{\psi}=\sum_{1\le n_1<\cdots<n_r\le N} a(n_1,\ldots,n_r)\ket{n_1,\ldots,n_r}$, with coefficients given by 
\begin{align} \label{coeffs}
a(n_1,\ldots,n_r)=\sum_{\mathcal P\in S_r}e^{i\sum_j k_{\mathcal P_j}n_j+\frac{i}{2}\sum_{i<j}\theta_{\mathcal P_i\mathcal P_j}},
\end{align}
where $S_r$ is the set of permutations of the labels $\{1,2,...,r\}$ and the parameters $k_i$ and $\theta_{ij}$ are the magnon momenta and phase angles, respectively \cite{BA1,BA2}. Within the coordinate Bethe ansatz, these parameters satisfy the following ``Bethe ansatz equations'':
$2\cot\frac{\theta_{ij}}{2}=\cot\frac{k_i}{2}-\cot\frac{k_j}{2}$ and $Nk_i=2\pi\lambda_i+\sum_{j\neq i}\theta_{ij}$ where $\{\lambda_i\}\in \{0,1,...,N-1\}$ are the Bethe quantum numbers. The eigenenergy is given by $E=\sum_{i=1}^{r}(\cos k_i-1)+\frac{JN}{4}$. At finite $N$, the ground-state energy is algebraic; it is a root of the characteristic polynomial of \cref{Ham}. It follows that the momenta and phase angles can always be written as $k_i, \theta_{ij} \rightarrow \arccos \alpha$, with algebraic $\alpha$ (see \cref{tab:E0}). This should be contrasted with the well-known thermodynamic-limit energy density of $-\ln 2+1/4$, which is transcendental \cite{hulthen1938uber}. The finite-size problem therefore sits in an interesting intermediate regime: the model is exactly solvable, the finite energies are algebraic, yet the algebraic structure of the exact roots is highly nontrivial. 

Albeit the coordinate Bethe ansatz allows for an intuitive and physical interpretation of the eigenstates in terms of magnon excitations, the coupled nonlinear equations can be entirely forgone in favour of a purely algebraic determination of the so called ``Bethe roots'' (also known as the rapidities) \cite{faddeev1996algebraicbetheansatzworks,Staudacher}, which are related to the magnon momenta and energy by $u_i = \frac{1}{2}\cot \frac{k_i}{2}$ and $E = -\frac{1}{2}\sum_{i} (\frac{1}{u_i^2 + 1/4} -1 )$. The Bethe roots are the zeroes of the Baxter polynomial \cite{BAXTER19731}, defined as 
\begin{align} \label{Baxter_Poly}
    Q(u) & = \prod_{i=1}^{r}(u - u_i)  = u^r + \sum_{a=0}^{r-1} c_{a} u^{a}, 
\end{align}
where, for example, $c_{r-1} = -\sum_{i=1}^{r} u_i $ and $r$ is the number of magnons. We may think of the coefficients $c_a$ (the ``$Q$-coefficients'') as objects that collectively encode information about the Bethe roots \cite{Bajnok_QSystems_2020,SciPostPhys.16.5.129}, whereby they can be handled without arbitrarily assigning them individual labels. Bethe roots that solve the Bethe ansatz equations must also obey Baxter's $TQ-$relation, defined as \mbox{$T(u)Q(u) = F(u)$} where  
\begin{equation}
    F(u) = \left(u + \frac{i}{2} \right)^N Q(u-i) + \left( u - \frac{i}{2} \right)^N Q(u+i), 
\end{equation}
and we require that $T(u)$ is a polynomial in the spectral parameter $u$ such that $F(u)\,\text{mod}\,Q(u) = 0$. The coefficients on the $u$ dependent terms of the remainder form a polynomial system in $\{ c_a\}$ \cite{ZRC_Granet}. The roots that yield physical solutions to \cref{Ham} can be found efficiently by solving this system, which was first noted in the advent of the rational $Q$-system approach developed by Marboe and Volin \cite{Marboe_2017,  Bajnok_QSystems_2020,SciPostPhys.16.5.129, SciPostPhys.16.4.113}. Although this zero-remainder condition of the $TQ$-relation yields a finite set of candidate $Q$-coefficients, some of which are potentially unphysical or associated with excited states \cite{Hao_2013}, for the present investigation we select only those unique solutions that yield the lowest real-valued energy and validate the roots against numerically determined values. 

\textit{Symbolic Ground State and the Valence-Bond Basis}.---
Let us define the singlet pair state as
\begin{align}
[i;j] = \frac{1}{\sqrt{2}} \left( \ket{\uparrow}_i\ket{\downarrow}_j - \ket{\downarrow}_i\ket{\uparrow}_j \right),
\end{align}
which is an elementary valence-bond (VB) state between qubits on sites $i$ and $j$. It is known that any singlet state can be written in the non-crossing VB basis \cite{rumer1932nachrichten,hulthen1938uber,TemperleyLieb1971,Saito_noncrossing,BEACH2006142}, which provides a compact representation of finite antiferromagnetic ground states. The exact ground states of the two- and four-site chains are especially simple:
\begin{align}
\ket{\mathrm{GS}}_2 = [1;2], \qquad \ket{\mathrm{GS}}_4 = \frac{[1;2][3;4]+[1;4][3;2]}{\sqrt{3}}.
\end{align}
For larger system sizes the ground state develops into the Liang-Doucot-Anderson Resonating-Valence-Bond (RVB) state \cite{PhysRevLett.61.365, doi:10.1143/JPSJ.58.1403}, involving a superposition of long distance VB configurations. This unequally-weighted, long-distanced RVB state is often used, and with much success, as a variational wavefunction to approximate the Heisenberg antiferromagnetic ground state in low-dimensional spin-1/2 systems \cite{Seki_2020,PhysRevB.39.140,doi:10.1143/JPSJ.55.323}. The symbolic Bethe ansatz solutions in the computational basis allow us to reconstruct the ground state in the VB basis \cite{supp}, thereby providing a direct connection between the algebraic Bethe solutions and the physical structure of the many-body wavefunction. Figure~\ref{Figs} shows the exact symbolic ground states of the two-, four-, \mbox{six-,} and eight-site chains in the non-crossing VB basis. Notably, the algebraic complexity of the Bethe roots is manifestly different to that of the ground state quantities. Furthermore, the ground-state energy for $N\leq10$ is found to have algebraic degree equal to the number of non-crossing VB basis states up to rotations and reflections \cite{hulthen1938uber,doi:10.1143/JPSJ.58.1403}. However, at $N=12$ we find that the algebraic degree of the ground-state energy is \textit{lower} than the dimension of the symmetrically nonequivalent non-crossing VB basis (see \cref{tab:Galois_data}). Therefore, $N=12$ marks the first chain size at which the characteristic polynomial in this basis is not the minimal polynomial, which hints that the fundamental algebraic structure of the Heisenberg chain ground state is independent of the VB representation. 

\begin{table}[t]
\caption{\label{tab:E0}Exact ground-state energies of the finite-sized and infinite Heisenberg chain. The ten- and twelve-site cosine phase polynomials, $P^{10}(x)$ and $P^{12}(x)$, as well as the minimal polynomials of the ground-state energies, $P^N(E_0)$, are given in the supplemental materials \cite{supp}.}
\begin{ruledtabular}
\begin{tabular}{c|c} 
N & $E_0/N$ (Exact) \\
\hline
    2 & $-\frac{3}{4}$ \\
    4 & $-\frac{2}{4}$ \\
    6 & $-\frac{1}{6}\left(1 + \frac{\sqrt{13}}{2} \right)$ \\
    8 & $-\frac{1}{8}\left[  \frac{1}{3} \left( 4 + 13\left[\frac{13}{2} (5 + 3 \sqrt{3}i)\right]^{
   -\frac{1}{3}}  + [\frac{13}{2} \left(5 + 3 \sqrt{3}i\right)]^{\frac{1}{3}} \right) \right]$ \\
    10 & $-\frac{1}{10}\left( \frac{7}{2} - 2[ P^{10}_{2}(x) + P^{10}_{4}(x) ]\right)$ \\
    12 & $-\frac{1}{12}\Big(3 - 2[P^{12}_{6}(x)  + P^{12}_{9}(x)  + P^{12}_{11}(x)]  \Big)$ \\
    $\vdots$ & $\vdots$   \\
    $\infty$ & $-\ln2 +\frac{1}{4}$
\end{tabular}
\end{ruledtabular}
\end{table}

\textit{Six-Site Chain}.---
The six-site chain ground state is parametrised by a single independent phase angle, $\theta_{13}$, satisfying the irreducible, integer polynomial 
\begin{equation} \label{N6_coord_poly}
    P^6(x) = -4x^2+6x+1, 
\end{equation}
where $x=\cos\theta_{13}$. We term such expressions ``cosine phase polynomials'', which are to be contrasted with the Baxter polynomials, \cref{Baxter_Poly}, which are not defined on the ring of integers. We further define the ``$Q$-coefficient polynomial'', denoted $P^N(c_{a})$ where $a=\frac{N}{2} - 2$, as the irreducible polynomial of the highest degree $Q$-coefficient of the Heisenberg chain ground state. For $N=6$ we find the $Q$-coefficient polynomial is given by 
\begin{align}
    P^6(c_1) = 48 c_1^2 - 40 c_1 -9.
\end{align}
The physical solution to \cref{N6_coord_poly} is
$\cos \theta_{13} = \frac{3-\sqrt{13}}{4} $ and the exact ground-state energy density is given in \cref{tab:E0}. Lastly, we define the minimal polynomials of the Heisenberg chain ground-state energy as $P^N(E_0)$, which for the $N=6$ chain is  
\begin{align}
    P^6(E_0) = 4 E_0^2 + 8 E_0 -9.
\end{align}
The lowest root of $P^N(E_0)$ is always the ground-state energy. The six-site cosine phase root, $\cos \theta_{13}$, is expressible in the radicals and shares the same algebraic degree as the $Q$-coefficients and the ground-state energy. The symbolic ground state in the VB basis may likewise be written explicitly in closed form, see \cref{N6_fig}. Its structure already departs from the simple nearest-neighbour RVB form of the four-site chain, requiring the inclusion of longer-ranged singlet coverings in the superposition \cite{PhysRevLett.61.365}.

\textit{Eight-Site Chain}.---
The situation changes markedly for the eight-site chain, where the intricate algebraic structure of the ground state begins to emerge. Here the ground state depends upon two independent phase angles, $\theta_{14}$ and $\theta_{23}$. Solving the Bethe ansatz equations yields a single cosine phase polynomial constraining both angles simultaneously, given by 
\begin{align} \label{P8_x}
    P^{8}(x) & = 64 x^6-64 x^5-80 x^4+96 x^3-56 x^2-100 x+5, 
\end{align}
where the third and second real roots (in ascending order) yield $P^{8}_{3}(x) = \cos \theta_{14}$ and $P^{8}_{2}(x) = \cos \theta_{23}$. In contrast, the Q-coefficient and ground-state energy polynomial are found to be 
\begin{equation}
    P^8(c_2) = 216 c_2^3-180 c_2^2-54 c_2+5,
\end{equation}
and
\begin{equation}
    P^8(E_0) = E_0^3+4 E_0^2+E_0-1,
\end{equation}
respectively. The Galois groups of these polynomials can be computed via the Fieker-Kl\"{u}ners algorithm \cite{Galois_2014} which is implemented in the MAGMA computer algebra system \cite{MAGMA}. We find that the Galois group of $P^8(x)$ is the group \mbox{$Z_2 \wr Z_3 \cong Z_2 \cross A_4$}, whereas the Galois groups of $P^8(c_2)$ and $P^8(E_0)$ are both found to be $Z_3$. As all three polynomial groups are solvable it follows that the cosine phase roots, Bethe roots, and ground-state energy are expressible in radicals, albeit with non-trivial nesting in the cosine phase case. Notably, $P^8(x)$ is a non-cyclotomic polynomial of degree six that is Galois solvable, which contrasts with the ground-state energy of algebraic degree three, as was first noted by Hulth\'en \cite{hulthen1938uber}. Although the $N=8$ cosine phase polynomial is solvable despite being greater than degree five, the bipartite Heisenberg chain ground state becomes generically Galois unsolvable for all chains larger than eight-sites; this result was first demonstrated via the rational $Q$-system by Jacobsen \textit{et al} \cite{Jacobsen2019Torus}. Moreover, the eight-site chain marks the point of separation between the algebraic complexity of the magnon phases and the ground state itself; whereas for the $N=4$ and $6$ cases they are the same. This would suggest that the fundamental basis governing the algebraic structure of the ground state is that of the symmetrically nonequivalent non-crossing VB states \cite{doi:10.1143/JPSJ.58.1403}; put differently, the characteristic polynomial in the non-crossing VB basis (reduced for symmetry) and the minimal polynomial of the ground-state energy have equal degree for $N=8$. However, we will later see that this feature is only incidental, and does not hold for $N>10$. 

\textit{Ten- and Twelve-Site Chains}.--- The ten- and twelve-site chains both possess ground states that are not solvable in the radicals; this can be seen in \cref{tab:E0}, where we are no longer able to write the energy density in algebraic closed form for these cases. The relevant polynomials for both systems as well as the symbolic ground states are given in the supplemental materials \cite{supp}. The ten-site case is also the first chain length where the ground-state wavefunction requires phase angles that do not correspond to magnon momenta; we refer to such phase angles as auxiliary as they are not described by the cosine phase polynomials $P^N(x)$. The ten-site cosine phase polynomial $P^{10}(x)$ is found to be twice the degree of the ground-state energy polynomial, similar to the eight-site case, whereas the twelve-site cosine phase polynomial $P^{12}(x)$ is triple the degree of the ground-state energy polynomial. This is also reflected in the Galois groups of the cosine phase polynomials, which are found to be $Z_2 \wr S_6$ for ten-sites and $S_3 \wr S_{10}$ for twelve sites. As such, we conjecture that the algebraic degree of the  energy and the cosine phase minimal polynomials of the Heisenberg chain ground state for even $N>2$ go as 
\begin{equation} \label{degree_conjecture}
    \deg[P^N(E_0)] = 
    \genfrac{(}{)}{0pt}{}{\frac{N -2}{2}}{\left\lfloor \frac{N - 2}{4}\right\rfloor},
\end{equation}
and 
\begin{equation}
    \deg[P^N(x)] = \left\lfloor \frac{N}{4} \right\rfloor \deg[P^N(E_0)], 
\end{equation}
respectively. Similarly, from the finite-size data in \cref{tab:Galois_data} we suggest that for $N>8$ the Galois group of the ground-state energy polynomial and the cosine phase polynomial is likely $S_{M}$ and $S_{D} \wr S_{M}$ respectively, where $M$ is given by \cref{degree_conjecture} and $D = \lfloor N/4 \rfloor$ is the number of distinct $Q$-coefficients. The factor of $\lfloor N/4 \rfloor$ arises from the inversion symmetry of the ground state, whereby the Bethe rapidities are invariant under reflection and occur in pairs (i.e. the $\frac{N}{2}$ magnon momenta need only be characterized by $\lfloor N/4 \rfloor$ non-trivial unique roots). Intuitively, the disparity between the degrees of $P^N(x)$ and $P^N(E_0)$ comes from the nature of what each object is describing. The ground-state energy is a collective quantity that does not care about the arbitrary labelling of the phases; it is invariant under permutations on the set of Bethe roots \cite{Yang_1966}. In contrast, an individual phase angle is only invariant up to reflection, and thus holds greater algebraic information to quantify the associated magnon.  

An important physical result from \cref{degree_conjecture} is that the underlying algebraic structure of the Heisenberg chain ground state is not governed by the symmetrically nonequivalent VB configurations. The $N=12$ Heisenberg chain ground state is spanned by twelve non-crossing VB states up to rotations and reflections, see Fig. 3 in the supplemental materials \cite{supp}. Even though the VB states are non-orthogonal they are linearly-independent, and thus the characteristic polynomial of this eigenspace will be degree twelve. As the ground-state energy is the root of a degree ten minimal polynomial, there are two roots unaccounted for; this hints at the existence of a ten-dimensional subspace that governs the algebraic structure of the $N=12$ ground state. 

\begin{table}[htb]
\caption{\label{tab:Galois_data} Algebraic degree and Galois group of the ground-state energy $(E_0)$ and cosine phase ($x = \cos \theta_{ip}$) minimal polynomials. Note that $P^{N}(x)$ does not describe the auxiliary phase angles, only those phases that are also magnon momenta; that is, $k_i = \theta_{ip}$ when $p=\frac{N}{2} - i + 1$. The degree and Galois group of the ground-state energy is found to be the same as that of the $Q$-coefficients.}
\begin{ruledtabular}
\begin{tabular}{c|c|c|c|c} 
N & deg[$P^{N}(E_0)$] & Gal[$P^{N}(E_0)$] & deg[$P^{N}(x)$] & Gal[$P^{N}(x)$] \\
\hline
    2 & 1 & \text{Trivial} & 1 & \text{Trivial} \\
    4 & 1 & \text{Trivial} & 1 & \text{Trivial} \\
    6 & 2 & $Z_2$ & 2 & $Z_2$ \\
    8 & 3 & $Z_3$ & 6 & $Z_2 \wr Z_3$ \\
    10 & 6 & $S_6$ & 12 & $Z_2 \wr S_6$ \\
    12 & 10 & $S_{10}$ & 30 & $S_3 \wr S_{10}$ \\
\end{tabular}
\end{ruledtabular}
\end{table}

\textit{Conclusion}.--- We have analysed the algebraic structure of finite-size Bethe ansatz solutions for the ground state of the Heisenberg chain. We find that the degree of the irreducible polynomials that govern the ground-state energy and $Q$-coefficients scale at different rates to the Bethe rapidities and magnon phases. We conjecture that the degree of the ground-state energy is given by \cref{degree_conjecture}, which is the integer sequence A001405 \cite{oeisA001405}. The $N=12$ ground state is found to be the first instance where the algebraic degree of the ground-state energy does not match the number of symmetrically nonequivalent, non-crossing VB configurations. We explicitly demonstrate the onset of Galois unsolvability for the ten- and twelve-site Heisenberg chain ground states, in complete agreement with previous literature \cite{Jacobsen2019Torus}. 

Despite the fact that the exact ground state of finite antiferromagnetic Heisenberg chains has been known to high accuracy for nearly a century, the symbolic ground state is rarely presented beyond the six-site chain. Our findings therefore extend the known symbolic Heisenberg chain ground-state wavefunctions and energies up to the ten- and twelve-site cases respectively. These symbolic solutions and the resulting lack of Galois solvability reinforces that formal integrability and elementary algebraic solvability are distinct notions, even within the paradigmatic Heisenberg chain. The present work also indicates that the minimal algebraic structure of the Heisenberg chain ground state for $N\geq 12$ is not entirely elucidated by the non-crossing valence bond states alone, and it remains an open question as to what the nature of this structure is and how it connects to the Bethe ansatz. If such a structure can be derived independent of the valence bond representation or the Bethe ansatz, it may open new avenues of interrogating the exact ground state of two- and three-dimensional antiferromagnetic clusters. In particular, this has implications for the long-ranged resonating valence bond ground state of finite-sized square and honeycomb lattice Heisenberg antiferromagnets, where there are currently no known analytic solutions.  

\textit{Acknowledgements.---} ORB thanks Y. Wang an anonymous reviewer for helpful comments. This work is supported by the MEXT Quantum Leap Flagship Program (MEXT Q-LEAP) under Grant No. JPMXS0118069605. 
\hphantom{\cite{KOUZOUDIS_1998,BARWINKEL2000,oeisA006840,oeisA002995,oeisA006082}}
\bibliography{refs}

\onecolumngrid
\pagebreak
\newpage

\section*{Supplemental materials for ``Algebraic structure of finite-size Bethe solutions in the Heisenberg chain"}

Here we show the derivation of the core results of the main text; the symbolic ground-state energies and wavefunctions of the six-, eight-, and ten-site antiferromagnetic Heisenberg chains is computed via the algebraic and coordinate Bethe ansatz. The symbolic ground-state energy of the twelve-site chain is derived along with a high-precision numerical evaluation of the ground-state wavefunction. We largely follow the notation of Ref. \cite{BA1} for the coordinate Bethe ansatz and the notation of Ref. \cite{Staudacher} for the algebraic Bethe ansatz, and we always take the roots of polynomials to be listed with the real roots first and in ascending order. Unless otherwise stated, all presented polynomials are irreducible. Computation of the Galois groups was performed with the MAGMA computer algebra system hosted by the University of Sydney \cite{MAGMA}. It will be useful later to briefly recall some results from the algebraic and coordinate Bethe ansatz (BA). For a state with $N$ sites and $r$ magnons the BA equations in terms of the magnon momenta ($k_i$) and phases ($\theta_{ij}$) are written as  
\begin{equation} \label{BA_phase_relation}
    2\cot\frac{\theta_{ij}}{2} = \cot\frac{k_i}{2} - \cot\frac{k_j}{2}, \qquad i,j = 1,...,r,
\end{equation}
and 
\begin{equation} \label{BA_momentum_relation}
    Nk_i = 2\pi \lambda_i + \sum_{j\neq i} \theta_{ij}, \qquad i=1,..,r, 
\end{equation}
with Bethe quantum numbers $\lambda_i \in \{0,1,...,N-1\}$. The Bethe roots are related to the magnon momenta via 
\begin{equation} \label{Bethe_roots_to_momenta}
    u_i = \frac{1}{2} \cot \frac{k_i}{2}.
\end{equation}
Defining the equation
\begin{equation}
    F(u) = \left(u + \frac{i}{2} \right)^N Q(u-i) + \left( u - \frac{i}{2} \right)^N Q(u+i),
\end{equation}
the condition for Baxter polynomiality is written as  
\begin{equation} \label{Bax_Poly_condition}
    F(u)\,\text{mod}\,Q(u) = 0,
\end{equation}
such that the remainder is a polynomial in $u$. The coefficients on the $u$ terms form a polynomial system over the $Q$-coefficients with a zero-dimensional ideal. These are not to be confused with the BA coefficients, which are the unnormalized coefficients on the computational basis states, and are given by 
\begin{align} \label{coeffs}
a(n_1,\ldots,n_r)=\sum_{\mathcal P\in S_r}e^{i\sum_j k_{\mathcal P_j}n_j+\frac{i}{2}\sum_{i<j}\theta_{\mathcal P_i\mathcal P_j}},
\end{align}
where $S_r$ is the set of permutations of the labels $\{1,2,...,r\}$. 

\section{Heisenberg chain ground state} \label{sec:HGS} 

The BA coefficients of the Heisenberg chain ground state are always real for even $N$. The even $N$ site AF ground state $\ket{\text{G.S.}}_N$ is a spin singlet and exists in the zero $z-$axis magnetization, $M^z = 0$, subspace such that $r = N/2$. The dimension of this subspace is the $r^{\text{th}}$ central binomial coefficient given by $\frac{(2r)!}{(r!)^2}$. The Bethe quantum numbers for the antiferromagnetic ground state are known to take the form $\{\lambda_i\} = \{1,3,5,...,N-1\}$. Since the ground state of the Heisenberg chain possesses inversion symmetry, the magnon momenta are evenly distributed about $k=\pi$. Furthermore, the Bethe roots are real and occur in opposite pairs about $u=0$, such that the Baxter polynomial of the Heisenberg chain ground state always takes the form 
\begin{align} \label{Bax_Poly_Supp}
    Q(u) & = (u^2 - b_1^2)(u^2 - b_2^2)(u^2 - b_3^2)... , \nonumber \\
    & = u^r + c_{r-2} u^{r-2} + c_{r-4} u^{r-4} ... \,\, .
\end{align}
The Bethe roots are therefore $\{ u_i\} = \{\pm b_1, \pm b_2, \pm b_3,... \}$ for $N/2$ even. Note that for $N/2$ odd we simply multiply \cref{Bax_Poly_Supp} by an extra factor of $u$ and add zero to the set $\{ u_i\}$. The highest degree $Q$-coefficient is therefore
\begin{equation}
    c_{r-2} = -\sum_i b_i^2. 
\end{equation}
Due to the inversion symmetry it follows that 
\begin{align} \label{symm_momenta}
    k_{1+i} &= 2\pi - k_{r-i}, \qquad i = 0,...,\floor*{\frac{r}{2}}, 
\end{align}
and from \cref{BA_phase_relation} and the fact that $\cot(2\pi - x) = -\cot(x)$, we find  
\begin{equation} \label{my_theta_conjecture}
    k_i = \theta_{ip}, \qquad p = r-i+1. 
\end{equation}
The momentum distribution also implies that for the case where $N/2$ is odd we recover the known results $k_{\ceil*{\frac{r}{2}}} = \pi$ and thus $\{ u_i\} = \{0,\pm b_1, \pm b_2, \pm b_3,... \}$.

\begin{figure}[htb] 
    \centering
    \begin{tikzpicture}[scale=0.8]

      \foreach \x [count=\xi from 1] in {1,2,3,4,5,6} {
        \foreach \y [count=\yi from 1] in {1, 2, 3, 4, 5, 6} {
        
            \ifnum\xi=\yi
                \fill[black] (\xi-2,5-\yi) rectangle ++(1,1);
            \fi
            \ifnum\yi>\xi
                \fill[yellow!50] (\xi-2,5-\yi) rectangle ++(1,1);
            \fi
            \ifnum\yi=1
                \ifnum\xi=6
                    \fill[cyan!50] (\xi-2,5-\yi) rectangle ++(1,1);
                \fi
            \fi
            \ifnum\yi=2
                \ifnum\xi=5
                    \fill[cyan!50] (\xi-2,5-\yi) rectangle ++(1,1);
                \fi
            \fi
            \ifnum\yi=3
                \ifnum\xi=4
                    \fill[cyan!50] (\xi-2,5-\yi) rectangle ++(1,1);
                \fi
            \fi
            \ifnum\yi>1
                \ifnum\xi=6
                    \ifnum\yi<6
                        \fill[orange!50] (\xi-2,5-\yi) rectangle ++(1,1);
                    \fi
                \fi
            \fi
            \ifnum\yi>2
                \ifnum\xi=5
                    \ifnum\yi<5
                        \fill[orange!50] (\xi-2,5-\yi) rectangle ++(1,1);
                    \fi
                \fi
            \fi
            \ifnum\xi>1
                \ifnum\yi=1
                    \ifnum\xi<6
                        \fill[green!50] (\xi-2,5-\yi) rectangle ++(1,1);
                    \fi
                \fi
            \fi
            \ifnum\xi>2
                \ifnum\yi=2
                    \ifnum\xi<5
                        \fill[green!50] (\xi-2,5-\yi) rectangle ++(1,1);
                    \fi
                \fi
            \fi
          \node at (\xi-1.5, 5.5-\yi) {$\theta_{\yi \xi}$};
        }
      }
      \draw[step=1cm] (-1,-1) grid (5,5);
    \end{tikzpicture}
    \caption{\label{Phases_diagram} The magnon phase angles of the $N=12$ Heisenberg chain ground state are antisymmetric about the diagonal and symmetric about the antidiagonal. Only the green and cyan phase angles are needed to fully characterise the ground state, and three of the green phase angles are constrained by the cyan phase angles.}
\end{figure}
 
As $\theta_{ij} = -\theta_{ji}$ and $\theta_{ii}=0$ by definition, the maximum number of phase angles that characterise an eigenstate of the Heisenberg chain is $\frac{r(r-1)}{2}$. However, the inversion symmetry of the ground state allows us to reduce this number via \cref{my_theta_conjecture}, such that there are always $\floor*{\frac{r}{2}}$ phases corresponding to magnon momenta, each one of which constrains one other magnon phase due to \cref{BA_momentum_relation,symm_momenta}. Moreover, we note that the phase angles themselves are also symmetric about the magnon momenta in the ground state, which leads to the relation
\begin{equation}
    \theta_{ab} = \theta_{cd}, \qquad c = r-b+1, \qquad d = r-a+1.
\end{equation}
It follows that the number of independent phase angles, $n_{\theta}(r)$, that parametrise the Heisenberg chain ground state with $r$ magnons is given by 
\begin{align}
    n_{\theta}(r) & = \left\lfloor \frac{r^2}{4} \right\rfloor - \left\lfloor \frac{r}{2} \right\rfloor \nonumber \\
    & = \left\lfloor \frac{(r-1)^2}{4} \right\rfloor
\end{align}
This is most easily seen through a diagrammatic example. Consider the $N=12$ chain with $r=6$ magnons. The phases are labelled by the indices of a $6\times6$ matrix, see \cref{Phases_diagram}, with the phases along the diagonal equal to zero. The lower triangular entries (yellow) are the negatives of the corresponding upper triangular entries, and are thus not independent. The upper antidiagonal entries (cyan) correspond to those magnon phases that are also magnon momenta; there are three independent momentum values, as the lower antidiagonal entries are again the negatives. All pairwise upper triangular entries about the antidiagonal are equal (e.g. $\theta_{24} = \theta_{35}$) such that their are only nine unique phase values; the six entries above the diagonal and antidiagonal (green) plus the three magnon momenta phases. However, \cref{BA_momentum_relation} constrains three of these nine values, and we can thus identify that the number of independent phase angles for the twelve-site ground state is six. These numbers are known as the `quarter square' numbers. 

\section{$N=6$ Heisenberg chain}

Here we derive the symbolic ground state of the six-site chain using the algebraic and coordinate Bethe ansatz. 
\subsection{Algebraic BA}

For the $N=6$ ground state the Baxter Polynomial is  
\begin{equation} \label{Bax_N6}
    Q(u) = u^3 + c_1 u .
\end{equation}
Baxter polynomiality yields the following single polynomial in $c_1$:
\begin{equation}
    0 = -\frac{1}{32} (4 c_1 -1) ( 48  c_1^2 - 40 c_1 -9 ).
\end{equation}
The quadratic factor contains the relevant ground state roots which are $c_1 = \frac{1}{12} (5 \pm 2 \sqrt{13})$, and is therefore the minimal $Q$-coefficient polynomial. It is important to note that the solution associated with the term $(4c_1 - 1)=0$ yields roots that are singular \cite{faddeev1996algebraicbetheansatzworks, Hao_2013}. That is, when $c_1 = \frac{1}{4}$ the non-trivial solutions to \cref{Bax_N6} are $u=\pm \frac{i}{2}$. Thus, the zero-remainder condition of the $TQ$-relation employed in this work does not immediately yield the complete physical spectrum, but rather a number of candidate solutions. We thus identify the physical, ground-state solution as that which has real, inversion symmetric roots and yields the lowest energy. 

Substituting the minus sign root (which is the ground-state solution) into \cref{Bax_N6} yields what we shall term the Bethe rapidity polynomial for the six-site Heisenberg chain ground state: 
\begin{equation} \label{P6u}
    P^6(u) = 48 u^4+40 u^2-9. 
\end{equation}
Note that this polynomial is even and explicitly twice the degree of the minimal $Q$-coefficient polynomial. We can directly derive the cosine phase polynomial, $P^6(x=\cos \theta_{13})$, by first noting that $\cos(\arccot x) = \frac{x}{\sqrt{1+x^2}}$, which combined with \cref{Bethe_roots_to_momenta} yields
\begin{equation}
    \cos k_i = \frac{4u_i^2 - 1}{4u_i^2 + 1},
\end{equation}
or alternatively
\begin{equation} \label{root_sub}
    u_i^2 = \frac{1+\cos k_i}{4(1 - \cos k_i)}. 
\end{equation}
Substituting into \cref{P6u}, clearing denominators, and defining $x = \cos k_1 = \cos \theta_{13}$ yields
\begin{equation} \label{P6_x}
    P^6(x) = -4 x^2 + 6 x + 1. 
\end{equation}
\subsection{Coordinate BA}
The Bethe quantum numbers for the $N=6$ chain ground state are $\{\lambda_i\} = \{1,3,5\}$. Consider the case of $\lambda_1 = 1$, which yields 
\begin{equation}
    6k_1 = 2\pi + \theta_{12} + \theta_{13}. 
\end{equation}
From \cref{my_theta_conjecture} we find the phase angles $\theta_{13}$ and $\theta_{12}$ are constrained by 
\begin{equation} \label{N6_theta12_constraint}
    \theta_{12} = 5\theta_{13} - 2\pi.
\end{equation}
As we already know $k_1 = \theta_{13}$ and $k_2 = \pi$ for $N=6$ via \cref{my_theta_conjecture}, we find that
\begin{align}
    2\cot(\frac{5\theta_{13}}{2}) & = \cot(\frac{\theta_{13}}{2}).
\end{align}
Expanding the cotangents into reduced trigonometric form yields 
\begin{align}
     0 & = 2 \cos(\frac{5 \theta_{13}}{2}) \sin(\frac{\theta_{13}}{2}) - \cos(\frac{\theta_{13}}{2})\sin(\frac{5 \theta_{13}}{2})  \nonumber \\
      & = -4 \cos^2(\theta_{13}) +6 \cos(\theta_{13}) + 1, 
\end{align}
which is exactly the polynomial given in \cref{P6_x} and the physical solution is $\cos \theta_{13} = \frac{3 - \sqrt{13}}{4}$. 
\subsection{Ground state quantities}
The ground-state energy of the six-site chain is given by 
\begin{align}
    E_{0} & = 2\cos \theta_{13} -\frac{5}{2} \nonumber \\
    & = -\left( 1 + \frac{\sqrt{13}}{2} \right). 
\end{align}
This result was first given by Hulth\'{e}n \cite{hulthen1938uber} in 1938 and then later by Kouzoudis \cite{KOUZOUDIS_1998} in 1998. The minimal polynomial of the ground-state energy is found to be
\begin{equation}
    P^6(E_0) = 4 E_0^2+8 E_0-9.
\end{equation}
Note that this polynomial has the same degree as the $Q$-coefficient polynomial, which we will later see is a pattern that holds for all chain lengths we consider. For all ground-state energy polynomials the lowest valued root is the ground-state energy in accordance with the Cayley-Hamilton theorem. As the momenta and phase factors can all be written in terms of only $\theta_{13}$, we may write the BA coefficients, \cref{coeffs}, in closed form as trigonometric functions in $\theta_{13}$. There are three unique (up to a minus sign) BA coefficients for the ground state of the six-site chain, namely
\begin{align}
    \abs{a(i,i+1,i+2)} & = 2\cos(\frac{7\theta_{13}}{2}) -4 \cos(\frac{\theta_{13}}{2})\nonumber \\
    & = \frac{1}{2}(41 - 11\sqrt{13}), \\
    \abs{a(i,i+1,i+3)} & = 2\cos(\frac{\theta_{13}}{2}) - 2\cos(\frac{3\theta_{13}}{2})  + 2\cos(\frac{5\theta_{13}}{2}) \nonumber \\
    & =  -5 + 2\sqrt{13},\\
    \abs{a(i,i+2,i+4)} & = -6 \cos(\frac{3\theta_{13}}{2})\nonumber \\
    & =  \frac{3}{2}(7 - \sqrt{13}),
\end{align}
where $i\in[1,..,6]$ and addition is taken as 1-based modulo 6. The multiplicities of these coefficients are 6, 12, and 2, respectively. This yields a normalization factor, $N_6$, of 
\begin{align}
    N_6 & = \sqrt{6\,|a(1,2,3)|^2 + 12\,|a(1,2,4)|^2 + 2\,|a(1,3,5)|^2} \nonumber \\
    & = 6\sqrt{169 - 46\sqrt{13}}.
\end{align}
Let $\sum^{\text{cyc}}_{i \in |\mathcal{G}|}$ denote the sum over the $|\mathcal{G}|$ non-equivalent cyclic permutations of the spin configuration of a basis state and L.C. is the local complement. For example, the state $\ket{\downarrow\uparrow\uparrow\uparrow}$ has $|\mathcal{G}| = 4$; we therefore may write
\begin{equation}
    \sum^{\text{cyc}}_{i \in 1,...,|\mathcal{G}|} \ket{\downarrow\uparrow\uparrow\uparrow} \equiv \ket{\downarrow\uparrow\uparrow\uparrow} + \ket{\uparrow\downarrow\uparrow\uparrow} + \ket{\uparrow\uparrow\downarrow\uparrow} + \ket{\uparrow\uparrow\uparrow\downarrow}, 
\end{equation}
and the L.C. of $\ket{\downarrow\uparrow\uparrow\uparrow}$ is $\ket{\uparrow\downarrow\downarrow\downarrow}$. The ground state of the six-site Heisenberg chain in the computational basis is therefore

\begin{align} \label{N6_GS_comp}
    \ket{\text{G.S.}}_{6} & = \sum^{\text{cyc}}_{i \in 1,...,6} (-1)^{i} \left[ -\frac{1}{6} \sqrt{\frac{1}{2} \left(5 - \frac{17}{ \sqrt{13}}\right)} \ket{\downarrow\downarrow\downarrow\uparrow\uparrow\uparrow}  + \frac{1}{6} \sqrt{1 + \frac{2}{\sqrt{13}}} \Big(\ket{\downarrow\downarrow\uparrow\downarrow\uparrow\uparrow} - \text{L.C.} \Big) \right] \nonumber \\
    & \kern+2em + \sqrt{\frac{1}{8} + \frac{3}{8 \sqrt{13}}} \left( \ket{\downarrow\uparrow\downarrow\uparrow\downarrow\uparrow} - \text{L.C.}\right),
\end{align}
and in the non-crossing valence-bond (VB) basis is 
\begin{align}
    \ket{\text{G.S.}}_{6} & = \sqrt{1-\frac{3}{\sqrt{13}}} \Bigg( [1;2][3;4][5;6] - [1;6][3;2][5;4] \Bigg) \nonumber \\
    & \kern+2em +  \frac{1}{3}\sqrt{5-\frac{17}{\sqrt{13}}} \Bigg(  [1;2][3;6][5;4]  + [1;4][3;2][5;6] + [1;6][3;4][5;2] \Bigg) .
\end{align}

\section{$N=8$ Heisenberg Chain}

The eight-site case is the first instance for which the ground-state energy and the magnon cosine phases possess different algebraic degrees. Notably, both quantities remain solvable in the radicals. 
\subsection{Algebraic BA}

For the $N=8$ ground state the Baxter Polynomial is  
\begin{equation} \label{Bax_N8}
    Q(u) = u^4 + c_2 u^2 +  c_0.
\end{equation}
The remainder of $F(u)\,\text{mod}\,Q(u)$ for the $N=8$ case, which we denote $r_8$, is given by 
\begin{align}
    r_8 & = \frac{1}{128} \Big[ (2560 c_0^2 - 4096 c_0 c_2^2 + 1280 c_0 c_2 - 448 c_0 + 768 c_2^4 - 1536 c_2^3 + 672 c_2^2 + 144 c_2 - 54)  u^2  \nonumber \\
    & \kern+2em -3328 c_0^2 c_2-256 c_0^2+768 c_0 c_2^3-1536 c_0 c_2^2+672 c_0 c_2+64 c_0-c_2+1 \Big].
\end{align}
Computing a Gr\"obner basis of the polynomial system formed by the two coefficients in $r_8$ and eliminating $c_2$ and $c_0$ yields the polynomials 
\begin{equation}
    0 = (2 c_2-1)^4 (24 c_2^3-68 c_2^2+10 c_2+1) (216 c_2^3-180 c_2^2-54 c_2+5),
\end{equation}
and 
\begin{equation}
    0 = (16 c_0-1)^4 (12288 c_0^3-6400 c_0^2-624 c_0-3) (552960 c_0^3-122112 c_0^2+336 c_0+1).
\end{equation}
The rightmost cubic factors are the minimal $Q$-coefficient polynomials:
\begin{subequations}
    \begin{align}
        P^8(c_2) & = 216 c_2^3-180 c_2^2-54 c_2+5, \\
        P^8(c_0) & = 552960 c_0^3-122112 c_0^2+336 c_0+1.  
    \end{align}
\end{subequations}
The physical solutions for the ground state are $P_1^8(c_2)$ and $P_2^8(c_0)$. Substituting these physical values of $c_2$ and $c_0$ into \cref{Bax_N8} and eliminating yields the eight-site Bethe rapidity polynomial of
\begin{equation}
    P^8(u) = 552960 u^{12}+460800 u^{10}-16128 u^8+81664 u^6-36464 u^4+648 u^2-1.
\end{equation}
We can once again solve for the cosine phase polynomial via substitution of \cref{root_sub} to find 
\begin{equation} \label{P8_x}
    P^8(x) = 64 x^6-64 x^5-80 x^4+96 x^3-56 x^2-100 x+5.
\end{equation}
Note the fact that the degree of the Bethe rapidity polynomial is twice the degree of the cosine phase polynomial and four times the degree of the $Q$-coefficient polynomials. The Galois group of $P^{8}(x)$ is found to be $Z_2 \wr Z_3 \cong C_2 \times A_4 $ using the MAGMA computer algebra system. This group is a solvable Galois group, most readily seen as it is of order 24, which is less than the smallest non-Abelian composition factor (the $A_5$ group) which has order 60. It follows that the roots of $P^8(x)$ are expressible in radicals, which means the $N=8$ ground state cosine phase polynomial is an example of a non-cyclotomic polynomial greater than degree five that is Galois solvable. $N=8$ is the largest size of the bipartite Heisenberg chain for which this is true, as was first noted by Jacobsen, Jiang and Zhang in 2018 \cite{Jacobsen2019Torus}. 

\subsection{Coordinate BA}
Noting that the Bethe quantum numbers for the eight-site case are $\{1,3,5,7\}$ and that $\theta_{14} = k_1$ and $\theta_{23} = k_2$, it follows from \cref{my_theta_conjecture,symm_momenta} that 
\begin{subequations}
\begin{align}
    7\theta_{14} - 2\pi & = \theta_{13} + \theta_{12}, \\
    7\theta_{23} - 6\pi & = \theta_{13} - \theta_{12}.  
\end{align}
\end{subequations}
Eliminating $\theta_{12}$ and $\theta_{13}$ via \cref{BA_phase_relation} yields
\begin{equation}
    2\cot(\frac{7}{4}[\theta_{14}\pm \theta_{23}]) = \cot(\frac{\theta_{14}}{2}) \pm \cot(\frac{\theta_{23}}{2}). 
\end{equation}
Again, expanding the cotangents yields 
\begin{align}
    0 & = -2 \cos(\frac{5 (\theta_{14}\pm\theta_{23})}{4})+ \cos(\frac{9 \theta_{14}\pm5 \theta_{23}}{4}) + \cos(\frac{5 \theta_{14}\pm9 \theta_{23}}{4}). \label{cosab}
\end{align}
Letting $a = \frac{1}{4}(\theta_{14} + \theta_{23})$ and $b = \frac{1}{4}(\theta_{14} - \theta_{23})$ the two equations in \cref{cosab} can equivalently be written
\begin{align}
    0 &= \cos(2a)\cos(7b) - \cos(5b), \\
    0 & = \cos(2b)\cos(7a) - \cos(5a). 
\end{align}
As these equations are symmetric we need only trigonometric expand one and self substitute to yield a single variable polynomial. The expanded trigonometric form yields 
\begin{align}
    0 & = (2\cos^2(a) - 1)\Big(-7\cos(b) + 56 \cos^3(b) -112\cos^5(b) + 64 \cos^7(b)\Big) - \Big(5 \cos(b)-20 \cos^3(b)+16 \cos^5(b)\Big). 
\end{align}
Letting $\cos a = z$, we eliminate to the polynomial 
\begin{align}
     0 & = (z-1) (z+1) (8 z^3-4 z^2-4 z+1)  (8 z^3+4 z^2-4 z-1) (4096 z^{12}-16384 z^{10}  +26368 z^8-22016 z^6 \nonumber \\
     & \kern+2em +10144 z^4 -2448 z^2+235). \label{poly_12}
\end{align}
The degree twelve polynomial factor contains the relevant roots of the ground state, as can be verified by numeric evaluation of the BA equations \cite{BA2}. Let us define the temporary polynomial $T^8(z)$ as
\begin{align} \label{P8_inz}
    T^8(z) & = 4096 z^{6}-16384 z^{5}+26368 z^4  -22016 z^3+10144 z^2-2448 z +235,
\end{align}
and the physical roots of the ground state are $T_1^8(z) = \cos^2 a$ and $T_3^8(z) = \cos^2 b$. We may re-parametrize the polynomial in terms of $\cos \theta_{14}$ and $\cos \theta_{23}$ by noting  
\begin{align}
        \cos \theta_{14} & = \cos\left[ 2\arccos(\sqrt{T^8_{1}}) - 2\arccos(\sqrt{T^8_{3}}) \right]\nonumber \\
        & = (2T^8_{1} - 1)(2T^8_{3} - 1)  + 4 T^8_{1}T^8_{3} \sqrt{1 - T^8_{1}} \sqrt{1 - T^8_{3}}. 
\end{align}
Computing the minimal polynomial yields the cosine phase polynomial given in \cref{P8_x}, where the physical roots ($ P^{8}_{3}(x) = \cos\theta_{14} $ and $ P^{8}_{2}(x) = \cos\theta_{23}$) are 
\begin{align}
    P^{8}_{3}(x) & = \frac{1}{312} \Bigg( 52 -13 \cdot 2^{\frac{2}{3}} (13 \xi)^{\frac{1}{3}} + 2^{-\frac{2}{3}} (-5 + 3 \sqrt{3} i ) (13 \xi)^{\frac{2}{3}} \nonumber\\
    & +  \sqrt{39} \sqrt{936 - 42\cdot 2^{\frac{1}{3}} (13 \xi)^{\frac{2}{3}} + 13\cdot 2^{\frac{2}{3}} (3 + 11 \sqrt{3} i )(13 \xi)^{\frac{1}{3}} + 16\cdot 2^{\frac{1}{3}}  (3 + \sqrt{3}i)) (-13 \xi)^{\frac{2}{3}} } \Bigg) \nonumber\\
   & = 0.0487744831019861... \,\,
\end{align}
and 
\begin{align}
    P^{8}_{2}(x) & = \frac{1}{312} \Bigg( 52 -13\cdot 2^{\frac{2}{3}} (13\xi)^{\frac{1}{3}} + 2^{-\frac{2}{3}} (-5 + 3 \sqrt{3} i ) (13 \xi)^{\frac{2}{3}} \nonumber \\
    & -  \sqrt{39} \sqrt{936 - 42\cdot 2^{\frac{1}{3}} (13\xi)^{\frac{2}{3}} + 13\cdot 2^{\frac{2}{3}} (3 + 11 \sqrt{3} i ) (13\xi)^{\frac{1}{3}}  + 16 \cdot 2^{\frac{1}{3}} (3 + \sqrt{3}i) (-13\xi)^{\frac{2}{3}}} \Bigg) \nonumber\\
    & = -0.8743211875705738... \,\, 
\end{align}
where $\xi = 5 + 3 \sqrt{3} i$. The values of these roots were verified via numerical evaluation of the BA equations. 

\subsection{Ground state quantities}
The $N=8$ ground-state energy is expressible in radicals and is given by 
\begin{align}
    E_0 & = 2\cos \theta_{14} + 2\cos \theta_{23} -2, \nonumber \\
    & = -\frac{1}{3}\left(  4 + 13\left[\frac{13}{2} (5 + 3 \sqrt{3}i)\right]^{
   -\frac{1}{3}} + \left[\frac{13}{2} \left(5 + 3 \sqrt{3}i\right)\right]^{\frac{1}{3}} \right).
\end{align}
This result was first implied by Hulth\'{e}n \cite{hulthen1938uber} in 1938 and then given explicitly by B\"{a}rwinkel, Schmidt, and Schnack in 2000 \cite{BARWINKEL2000}. The eight-site ground-state energy is an example of the \textit{casus irreducibilis}, as it is a real number that requires radicals of complex numbers. The minimal polynomial of the eight-site ground-state energy is given by 
\begin{equation}
    P^8(E_0) = E_0^3+4 E_0^2+E_0-1, 
\end{equation}
with a Galois group of $Z_3$. Again, the energy has the same algebraic degree as the $Q$-coefficients. There are seven unique BA coefficients of the eight-site ground-state wavefunction. For brevity, we introduce a small abuse of notation where we take $i=1$ when writing a coefficient such as $a(i,i+1,i+2,i+3)$, with the understanding that the cyclic permutations of such a configuration are equal. The coefficients in terms of the phase angles $\theta_{14}$ and $\theta_{23}$ are 
\begin{widetext}
\begin{align}
    a(1,2,3,4) & = 4 \cos(\frac{9\theta_{14}}{2} ) \cos(\frac{\theta_{23}}{2} )  + 8 \cos(2 \theta_{14}) \cos(2 \theta_{23}) + 4\cos(\frac{\theta_{14}}{2} ) \left(2 \cos(\frac{\theta_{23}}{2}) +  \cos(\frac{9\theta_{23}}{2} )\right)\nonumber \\
    a(1,2,3,5) & = 4\cos(\frac{\theta_{23}}{2} ) \left(  \cos(\frac{3\theta_{14}}{2} ) + \cos(\frac{7\theta_{14}}{2} ) \right)   + 
 4 \cos(2 \theta_{14}) \cos( \theta_{23}) + 4 \cos( \theta_{14}) \cos(2 \theta_{23}) \nonumber \\
 & \kern+2em + 4\cos(\frac{\theta_{14}}{2} ) \left( \cos(\frac{3\theta_{23}}{2} ) + \cos(\frac{7\theta_{23}}{2}  ) \right) \nonumber \\
    a(1,2,4,5) & = 8 \cos(\frac{\theta_{14}}{2}) \cos(\frac{\theta_{23}}{2}) + 8 \cos(\theta_{14}) \cos(\theta_{23}) + 
 4 \cos(\frac{7 \theta_{14}}{2}) \cos(\frac{3 \theta_{23}}{2}) + 
 4 \cos(\frac{3 \theta_{14}}{2}) \cos(\frac{7 \theta_{23}}{2})\nonumber \\
    a(1,2,4,6) & = 4 \cos(\theta_{14}) + 4 \cos(\theta_{23}) + \left(4 \cos(\frac{\theta_{14}}{2}) + 4 \cos(\frac{5 \theta_{14}}{2}) \right) \cos(\frac{3 \theta_{23}}{2}) \nonumber \\
    & \kern+2em + \cos(\frac{3 \theta_{14}}{2}) \left(4 \cos(\frac{\theta_{23}}{2}) + 4 \cos(\frac{5 \theta_{23}}{2}) \right)\nonumber \\
    a(1,2,4,7) & = 4 \cos(\frac{5\theta_{14}}{2} ) \cos(\frac{\theta_{23}}{2}  ) + 
 8 \cos( \theta_{14}) \cos(\theta_{23}) + 
 8 \cos(\frac{3\theta_{14}}{2}  ) \cos(\frac{3\theta_{23}}{2} ) + 
 4 \cos(\frac{\theta_{14}}{2}  ) \cos(\frac{5\theta_{23}}{2}  )\nonumber \\
    a(1,2,5,6) & = 8 + 8 \cos(\frac{\theta_{14}}{2}) \cos(\frac{\theta_{23}}{2}) + 
 8 \cos(\frac{5 \theta_{14}}{2}) \cos(\frac{5 \theta_{23}}{2}) \nonumber \\
    a(1,3,5,7) & = 8 + 8 \cos(\frac{3 (\theta_{14} - \theta_{23})}{2}) + 8 \cos(\frac{3 (\theta_{14} + \theta_{23})}{2}). \nonumber
\end{align}

\begin{table}[t]

\caption{\label{Tab_N8} Numeric values of the ground state Bethe ansatz coefficients of the $N=8$ antiferromagnetic Heisenberg chain and their multiplicities.  }
\begin{tabular}{c|c|c} 
Coefficient & Numeric value & Multiplicity \\
\hline
    $a(1,2,3,4)$ & 0.2851453663081126 & 8 \\
    $a(1,2,3,5)$ & -1.3262377338246392 & 16 \\
    $a(1,2,4,5)$ & 2.0821847350330531 & 16 \\
    $a(1,2,4,6)$ & -6.2760132284441398 & 16\\
    $a(1,2,4,7)$ & 5.2349208609276133 & 8 \\
    $a(1,2,5,6)$ & 3.4378814922026728 & 4 \\
    $a(1,3,5,7)$ & 15.2045019245375580 & 2
\end{tabular}
\end{table}

In closed form, these coefficients are given by 
\begin{align}
    a(1,2,3,4) & = 35 - 217 \left(\frac{13}{2\sqrt{3}}\right)^{2/3} ( 1 + \sqrt{3}i)(
 38835 + 5351\sqrt{3}i)^{-1/3} - \frac{1}{2}\left(\frac{13}{18}\right)^{1/3}(1 - \sqrt{3}i) (38835 + 5351\sqrt{3}i)^{1/3} \nonumber \\
    a(1,2,3,5) & =  \frac{1}{3}\left(-76 + 217 \left(13\sqrt{2} \right)^{2/3} \left( -18149 + 8205  \sqrt{3}i \right)^{-1/3} + \left(\frac{13}{2}\right)^{1/3}\left( -18149 + 8205  \sqrt{3}i \right)^{1/3} \right) \nonumber \\
    a(1,2,4,5) & = -\frac{1}{3}\left( 58 + 62\left(\frac{13}{\sqrt{2}} \right)^{2/3}(1-\sqrt{3}i)\left( -1175+237\sqrt{3}i \right)^{-1/3} + \left(2\sqrt{13}\right)^{2/3}(1+\sqrt{3}i)\left( -1175+237\sqrt{3}i \right)^{1/3} \right)\nonumber \\
    a(1,2,4,6) & = \frac{1}{3}\left( 32 - 91 \left( \frac{13}{2}\right)^{2/3} (1+\sqrt{3}i)\left( 3775 + 2883\sqrt{3} i \right)^{-1/3} - \frac{1}{2} \left(\frac{13}{2}\right)^{1/3} (1 - \sqrt{3}i)\left( 3775+2883\sqrt{3}i\right)^{1/3}\right)\nonumber \\
    a(1,2,4,7) & = -1 + 7\left( 13\sqrt{\frac{2}{3}}\right)^{2/3}(-225 + 31 \sqrt{3} i)^{-1/3} + \left( \frac{13}{18}\right)^{1/3}(-225 + 31 \sqrt{3} i)^{1/3}\nonumber \\
    a(1,2,5,6) & = \frac{1}{3}\left(160 - 619 \left(13\sqrt{2} \right)^{2/3}(1-\sqrt{3}i)(106735 + 17709 \sqrt{3}i)^{-1/3} - \left(\frac{13}{2} \right)^{1/3}(1+\sqrt{3}i) (106735 + 17709 \sqrt{3}i)^{1/3} \right)\nonumber \\
    a(1,3,5,7) & = \frac{4}{3}\left(22 - 7 \left(\frac{13}{2}\right)^{2/3}(1+\sqrt{3}i)\left( 131+15\sqrt{3}i \right)^{-1/3} - \frac{1}{2}\left( \frac{13}{2}\right)^{1/3} (1-\sqrt{3}i)\left(131+15\sqrt{3}i \right)^{1/3} \right)\nonumber
\end{align}

The numeric value of the coefficients and their multiplicities are given in \cref{Tab_N8}. It is important to note that \textit{before} normalization these coefficients are algebraic degree three. The normalization factor, $N_8$, is given by 
\begin{align}
    N_8 & = \sqrt{\substack{8\,\abs{a(1,2,3,4)}^2 + 16\,\abs{a(1,2,3,5)}^2  + 16\,\abs{a(1,2,4,5)}^2 + 16\,\abs{a(1,2,4,6)}^2 \\ + 8\,\abs{a(1,2,4,7)}^2 + 4\,\abs{a(1,2,5,6)}^2 + 2\,\abs{a(1,3,5,7)}^2}}, \nonumber \\
    & = 38.1738048669707678... \,, 
\end{align}
which in principle can be written as a symbolic expression in terms of only $\theta_{13}$ and $\theta_{24}$, but we do not include it here due to its cumbersome size. In closed form over the radicals it is given by 
\begin{equation}
    N_8 = (1+i)\sqrt{2}\sqrt{ 3^{4/3} 26^{2/3} (\sqrt{3}+i) \xi^{1/3} + \left( 3^{1/6} 26^{1/3} 918294 (\sqrt{3}+i) \xi^{-1/3} -30732 \right) i}
\end{equation}
where $\xi = 57414051 + 2071655  \sqrt{3} i$. The coefficients in the non-crossing VB basis can be deduced by expanding the VB configurations and gathering like terms. In doing so, we find that 
\begin{subequations}
\begin{align}
    \alpha & = \frac{4}{N_8} \Big[ a(1,2,5,6) - 2a(1,2,3,4) \Big],\nonumber \\
    & = \frac{1}{3}\sqrt{\frac{1}{5} \left(28 + 532\left(\frac{2}{13}\right)^{1/3}(1 -  \sqrt{3} i)\left( 10993 + 705 \sqrt{3}i\right)^{-1/3} - 2\left(\frac{2}{13}\right)^{2/3} (1+\sqrt{3}i)\left( 10993 + 705 \sqrt{3}i\right)^{1/3}\right)} \nonumber \\
    & = 0.3004773319903022...\,\, , \\
    \beta & =\frac{4}{N_8} a(1,2,3,4), \nonumber \\
    & = \frac{1}{6}\sqrt{\frac{1}{65} \left( 416 -   794 \left(13\sqrt{2}\right)^{2/3} (1-\sqrt{3}i) \left( 38011 + 24555 \sqrt{3} i \right)^{-1/3} - \left(2\sqrt{13}\right)^{2/3}( 1 + \sqrt{3}i) \left( 38011 + 24555 \sqrt{3} i \right)^{1/3} \right)} \nonumber \\
    & = 0.0298786424147968...\,\, ,  \\
    \delta & = \frac{2}{N_8} a(1,2,4,5) \nonumber \\
    & = \frac{1}{6}  \sqrt{\frac{1}{65} \left( 104 - 14\left( 13\sqrt{2} \right)^{2/3} (1 + \sqrt{3}i) \left(-131 + 15 \sqrt{3} i \right)^{-1/3} - \left( 2\sqrt{13}\right)^{2/3} (1-\sqrt{3}i)\left(-131 + 15 \sqrt{3} i \right)^{1/3} \right)} \nonumber \\
    & = 0.1090897143886554...\,\,.
\end{align}
\end{subequations}
These values validate the previous results found via exact diagonalization in the VB basis \cite{doi:10.1143/JPSJ.58.1403}. Note that the wave-function coefficients in either the computational or VB basis become algebraic degree six only \textit{after} normalization, which contrasts with the unnormalized BA coefficients which are degree three. The ground state of the eight-site antiferromagnetic Heisenberg chain in the computational basis is therefore
    \begin{align} \label{N8_GS_comp}
        \ket{\text{G.S.}}_{8} & =  \frac{1}{N_8}\Bigg[\sum^{\text{cyc}}_{i \in 1,...,8 } \Big[ a(1,2,3,4) \ket{\downarrow\downarrow\downarrow\downarrow\uparrow\uparrow\uparrow\uparrow} +  a(1,2,3,5) \Big(\ket{\downarrow\downarrow\downarrow\uparrow\downarrow\uparrow\uparrow\uparrow} + \text{L.C.}\Big) +  a(1,2,4,5) \Big(\ket{\downarrow\downarrow\uparrow\downarrow\downarrow\uparrow\uparrow\uparrow} + \text{L.C.}\Big) \nonumber \\
        & \kern+2em + a(1,2,4,6)\Big(\ket{\downarrow\downarrow\uparrow\downarrow\uparrow\downarrow\uparrow\uparrow} + \text{L.C.}\Big) + a(1,2,4,7)\ket{\downarrow\downarrow\uparrow\downarrow\uparrow\uparrow\downarrow\uparrow} \Big] + \sum^{\text{cyc}}_{i \in 1,...,4 } a(1,2,5,6) \ket{\downarrow\downarrow\uparrow\uparrow\downarrow\downarrow\uparrow\uparrow} \nonumber \\
        & \kern+4em + a(1,3,5,7)\Big(\ket{\downarrow\uparrow\downarrow\uparrow\downarrow\uparrow\downarrow\uparrow} + \text{L.C.}\Big) \Bigg],
    \end{align}
and in the non-crossing VB basis is
\begin{align}
    \ket{\text{G.S.}}_{8} & =  \alpha \Big( [1;2][3;4][5;6][7;8] + [1;8][3;2][5;4][7;6]\Big)  \nonumber \\
    & \kern+2em + \beta \Big( [1;2][3;8][5;6][7;4] + [1;4][3;2][5;8][7;6] + [1;6][3;4][5;2][7;8] + [1;8][3;6][5;4][7;2]  \Big) \nonumber \\
    & \kern+6em + \delta \Big( [1;2][3;8][5;4][7;6] + [1;4][3;2][5;6][7;8] + [1;8][3;4][5;2][7;6] + [1;2][3;6][5;4][7;8]  \nonumber \\
    & \kern+8em + [1;8][3;2][5;6][7;4] + [1;2][3;4][5;8][7;6] + [1;6][3;2][5;4][7;8] + [1;8][3;4][5;6][7;2]  \Big). 
\end{align}

\section{$N=10$ Heisenberg chain}

Here we compute the exact ground state of the ten-site Heisenberg chain. Notably, although the ground state is analytic we show that both the Bethe roots and the ground state itself do not admit an elementary algebraic closed form as the parameters are algebraic numbers with unsolvable Galois groups. 
\subsection{Algebraic BA}
For the $N=10$ ground state the Baxter Polynomial is  
\begin{equation} \label{Bax_N10}
    Q(u) = u^5 + c_3 u^3 +  c_1 u.
\end{equation}
Baxter polynomiality once again yields a remainder, $r_{10}$, with two terms in $u$ (i.e. one term dependent on $u^2$ and one on $u^0$): 
\begin{align}
    r_{10} & = \frac{1}{512} \Big[ \Big(-23552 c_1^2 c_3+7680 c_1^2+19456 c_1 c_3^3-25600 c_1 c_3^2+6784 c_1 c_3-960 c_1-3072 c_3^5+8960 c_3^4-7808 c_3^3 \nonumber \\
    & \kern+2em +864 c_3^2+1316 c_3-250 \Big)u^2 -10240 c_1^3 + 16384  c_1^2  c_3^2 - 16640  c_1^2  c_3 - 1024 c_1^2 - 3072 c_1 c_3^4 + 8960 c_1 c_3^3 \nonumber \\
    & \kern+4em - 7808 c_1 c_3^2 + 864 c_1 c_3 + 856 c_1  - 17 c_3 + 15\Big].
\end{align}
We compute the Gr\"obner basis of the polynomial system defined by the coefficients and eliminate variables to find
\begin{equation}
    0 = (2 c_3-1)^5 (48 c_3^4-192 c_3^3+168 c_3^2-5) (5184 c_3^6-39744 c_3^5+62640 c_3^4-21984 c_3^3-10852 c_3^2+4780 c_3+125),
\end{equation}
and 
\begin{align}
    0 & = (16 c_1-1)^5 (196608 c_1^4-147456 c_1^3+10752 c_1^2+3264 c_1+43) (6794772480 c_1^6-29047652352 c_1^5-13762363392 c_1^4 \nonumber \\
    & \kern+2em +3699032064 c_1^3+515123968 c_1^2-1583456 c_1-133715).
\end{align}
Numerical evaluation of the BA equations confirms the degree six factors contain the relevant roots of the ground state. We can thus identify the $N=10$ $Q$-coefficient polynomials:
\begin{subequations}
    \begin{align}
        P^{10}(c_3) & = 5184 c_3^6-39744 c_3^5+62640 c_3^4-21984 c_3^3-10852 c_3^2+4780 c_3+125, \\
        P^{10}(c_1) & = 6794772480 c_1^6-29047652352 c_1^5-13762363392 c_1^4  +3699032064 c_1^3+515123968 c_1^2-1583456 c_1-133715).  
    \end{align}
\end{subequations}
The physical solutions for the ground state are $P_1^{10}(c_3)$ and $P_4^{10}(c_1)$. The Bethe rapidity polynomial is found to be
\begin{align}
    P^{10}(u) & = 6794772480 u^{24}+52093255680 u^{22}+111151153152 u^{20}+73292316672 u^{18}-17715494912 u^{16} \nonumber \\
    & \kern+2em  -28123725824 u^{14}-5722226688 u^{12}-5079916544 u^{10}+27233024 u^8+2016169216 u^6-323599584 u^4 \nonumber \\
    & \kern +4em +13604560 u^2-133715, 
\end{align}
and the cosine phase polynomial is given by 
\begin{align} \label{P10_x}
    P^{10}(x) & = 12288 x^{12}-92160 x^{11}+264192 x^{10}-321024 x^9+14080 x^8+394112 x^7-388224 x^6 +32000 x^5+174368 x^4 \nonumber \\
    & \kern+2em -92680 x^3-8412 x^2+12692 x-1637. 
\end{align}
The Galois group of $P^{10}(x)$ is found to be the group $Z_2 \wr S_6$, which has order $12!! = 2^6 \cdot 6! = 46080$. As the full symmetric group $S_6$ is not a solvable group the cosine phases of the $N=10$ Heisenberg chain cannot be expressed in the radicals. From \cref{root_sub} it is clear that the algebraic degree of the rapidity and cosine phase polynomials obey 
\begin{equation}
    \deg[P^N(u)] = 2\deg[P^N(x)],
\end{equation}
which is once again observed here. It is somewhat less clear what is the pattern between the degree of the cosine phase polynomials and the ground state quantities; here, we note that for $N=8$ and $N=10$ there are two distinct $Q$-coefficients and that $\deg[P^N(x)]$ was double that of the $Q$-coefficient degree. 
\subsection{Coordinate BA}

The Bethe quantum numbers for the ten-site chain are $\{1,3,5,7,9\}$ and we note that $k_1 = \theta_{15}$ and $k_2 = \theta_{24}$. Although the ground-state energy can be entirely parametrized in terms of only two phase angles, similar to the eight-site case, $N=10$ is the first chain length that requires auxiliary phase angles (see sec. \ref{sec:HGS}), which we choose to be $\theta_{12}$ and $\theta_{14}$. All four angles are required to fully characterize the ground-state wavefunction. Similar manipulations of the BA equations as per the ten-site case yields four coupled trigonometric equations: 
\begin{subequations}
\begin{align}
    0 & = 3 \sin(\frac{\theta_{12} + \theta_{14}}{2} - 4\theta_{15}) - \sin(\frac{\theta_{12}+\theta_{14}}{2} - 5\theta_{15}) \\
    0 & = 3 \sin(\frac{\theta_{12} - \theta_{14}}{2} + 4\theta_{24}) - \sin(\frac{\theta_{12}-\theta_{14}}{2} + 5\theta_{24}) \\
    0 & = \cos(\frac{\theta_{12}}{2} - \frac{(\theta_{15} - \theta_{24})}{2}) - \cos(\frac{\theta_{12}}{2})\cos(\frac{\theta_{15}+\theta_{24}}{2}) \\
    0 & = \cos(\frac{\theta_{14}}{2} - \frac{(\theta_{15} + \theta_{24})}{2}) - \cos(\frac{\theta_{14}}{2})\cos(\frac{\theta_{15}-\theta_{24}}{2}).
\end{align}
\end{subequations}

To reduce these coupled equations to a polynomial system we employ the tangent half-angle (Weierstrass) substitution. Letting $a = \tan(\frac{\theta_{15}}{2})$, $b = \tan(\frac{\theta_{24}}{2})$, $c= \tan(\frac{\theta_{12}}{2})$, and $d = \tan(\frac{\theta_{14}}{2})$, the above equations become 
\begin{subequations}
\begin{align}
0 & = 2 c + 2 d - 14 a + 14 c d a - 36 c a^2 - 36 d a^2 + 24 a^3 - 24 c d a^3 - 84 c a^4 - 84 d a^4 + 252 a^5 - 252 c d a^5 + 336 c a^6 \nonumber \\
& \kern+2em + 336 d a^6 - 264 a^7 + 264 c d a^7 - 126 c a^8 - 126 d a^8 + 34 a^9 - 34 c d a^9 + 4 c a^{10} + 4 d a^{10} \\
0 & = 2 c - 2 d + 14 b + 14 c d b - 36 c b^2 + 36 d b^2 - 24 b^3 - 24 c d b^3 - 84 c b^4 + 84 d b^4 - 252 b^5 - 252 c d b^5 + 336 c b^6 \nonumber \\
& \kern+2em - 336 d b^6 + 264 b^7 + 264 c d b^7 - 126 c b^8 + 126 d b^8 - 34 b^9 - 34 c d b^9 + 4 c b^{10} - 4 d b^{10} \\
0 & = c a - c b + 2 a b \\
0 & = d a + d b - 2 a b.
\end{align}
\end{subequations}
Once again computing a Gr\"{o}bner basis and eliminating to the variable $a$ yields
\begin{align}
0 & = a^3 \left(-7 + 12 a^2 + 126 a^4 - 132 a^6 + 17 a^8 \right) (-405 - 12420  a^2 - 106002  a^4 - 279588  a^6 + 270317  a^8  + 1716536  a^{10} \nonumber \\
& \kern+1em + 1397028  a^{12} + 4960856  a^{14} - 106379  a^{16} - 31502644  a^{18} + 20224974  a^{20} - 3401140  a^{22} + 133715  a^{24})
\end{align}
The degree twenty-four polynomial factor contains the relevant roots that characterize the ground state, as confirmed via numerical evaluation of the BA equations. Although the phases were parametrized via a tangent half-angle substitution we may re-parametrize to the cosine form by noting that
\begin{align}
    \cos[2\arctan(a)] = \frac{2}{1+a^2} - 1,
\end{align}
and therefore
\begin{equation}
    a^2 = \frac{1 - \cos \theta_{ij}}{1 + \cos \theta_{ij}}.
\end{equation}
As usual, defining $x = \cos \theta_{ij}$ and substituting yields the $N=10$ cosine phase polynomial $P^{10}(x)$ given in \cref{P10_x} with 
\begin{align}
    P^{10}_{4}(x) & = \cos \theta_{15}  \nonumber \\
    & = 0.1772365105015484... \,\, ,
\end{align}
and 
\begin{align}
    P^{10}_{2}(x) & = \cos \theta_{24}  \nonumber \\
    & = -0.6836491333339734... \,\, ,
\end{align}
once again confirmed via numerical evaluation of the BA equations. 

\subsection{Ground state quantities}

The ground-state energy of the ten-site chain is given by 
\begin{equation}
    E_0 = 2\cos \theta_{15} + 2\cos \theta_{24} -\frac{7}{2}. 
\end{equation}
The minimal polynomial of the $N=10$ ground-state energy is 
\begin{align}
    P^{10}(E_0) = 64 E_0^6+384 E_0^5+144 E_0^4-1728 E_0^3-1940 E_0^2+408 E_0+795, 
\end{align}
for which the Galois group is $S_6$. Thus, this is the first ground-state energy of the Heisenberg chain with even $N$ that is not expressible in the radicals.
\begin{table}[ht]
\caption{\label{N10_table} Numeric values of the unique ground state Bethe ansatz coefficients (up to a minus sign) of the $N=10$ antiferromagnetic Heisenberg chain and their multiplicities.  }
\begin{tabular}{c|c|c} 
Coefficient & Numeric value & Multiplicity \\
\hline
    $|a(1,2,3,4,5)|$ & 0.0742478017091568 & 10 \\
    $|a(1,2,3,4,6)|$ & 0.4466336681203956 & 20 \\
    $|a(1,2,3,4,7)|$ & 0.9818522866090206 & 40 \\
    $|a(1,2,3,5,7)|$ & 3.7404603572732196 & 20\\
    $|a(1,2,3,5,8)|$  & 4.6799086310431422 & 40 \\ 
    $|a(1,2,3,5,9)|$  & 2.4421820302084956 & 10 \\ 
    $|a(1,2,3,6,7)|$  & 2.7586080706641990 & 20 \\ 
    $|a(1,2,3,6,8)|$  & 6.9918830335869456 & 20 \\ 
    $|a(1,2,4,5,8)|$  & 7.4703604905773999 & 10 \\ 
    $|a(1,2,4,6,8)|$  & 20.6370145365441257 & 20 \\ 
    $|a(1,2,4,6,9)|$  & 15.9571059055009835 & 20 \\ 
    $|a(1,2,4,7,8)|$  & 9.9470751585230585 & 20 \\ 
    $|a(1,3,5,7,9)|$  & 51.1971318178431861 & 2\\ 
\end{tabular}
\end{table}
There are thirteen unique (up to a minus sign) BA coefficients of the ten-site chain, parametrized by the four aforementioned phase angles, $\{ \theta_{15}, \theta_{24}, \theta_{12}, \theta_{14}\}$. The closed form expressions of the BA coefficients in terms of the phase angles are very large and are available upon reasonable request. Moreover, determination of the minimal polynomials and resulting Galois groups of the BA coefficients becomes computationally expensive with the inclusion of auxiliary phase angles. For example, the simplest BA coefficient corresponds to the Ne\'el ordered basis state, $a(1,3,5,7,9)$. Using a high-precision lattice basis reduction approximation we find the BA coefficient $a(1,3,5,7,9)$ is the fifth root of the polynomial 
\begin{align}
    0 & = 27 x^{12}-41196288375 x^{10}+14368161049551250 x^8-1245186235698653328125 x^6 \nonumber \\
    & \kern+2em +18625521631912551985937500 x^4-69487496974195966283203125000 x^2 \nonumber \\
    &\kern+4em +75922590384074279119110107421875. 
\end{align}
This polynomial possesses the Galois group $Z_2 \wr S_6$, which is notably the same as the cosine phase polynomial, $P^{10}(x)$. The minimal polynomials of the other coefficients are omitted for brevity but are available upon reasonable request; all were found to possess the Galois group $Z_2 \wr S_6$ and are therefore unsolvable. \\

The numeric values and multiplicities of the BA coefficients are given in \cref{N10_table}. The minimal polynomial of the ten-site auxiliary phase angles, which we denote $P^{10,\text{aux}}(x)$, that defines the values of the cosine phase angles $\cos \theta_{12}$ and $\cos \theta_{14}$ can be derived in a similar fashion to those above. It is found to be
\begin{align}
    P^{10,\text{aux}}(x) & = 243 x^{12}-67392 x^{11}+1036222 x^{10}-9943488 x^9+55998829 x^8-163700096 x^7+423939172 x^6 \nonumber \\
    & \kern+2em -421880960 x^5 +148668829 x^4+246013120 x^3-348625538 x^2-67967424 x+2159203,
\end{align}
with $P_{1}^{10,\text{aux}}(x) = \cos \theta_{12}$ and $P_{2}^{10,\text{aux}}(x) =\cos \theta_{14}$. The $N=10$ Heisenberg chain normalization factor is therefore
\begin{align}
    N_{10} & = \sqrt{\substack{ 10|a(1,2,3,4,5)|^2 + 20|a(1,2,3,4,6)|^2 + 40|a(1,2,3,4,7)|^2 + 20|a(1,2,3,5,7)|^2 + 40|a(1,2,3,5,8)|^2 \\  + 10|a(1,2,3,5,9)|^2 + 20|a(1,2,3,6,7)|^2 + 20|a(1,2,3,6,8)|^2 + 10|a(1,2,4,5,8)|^2 + 20|a(1,2,4,6,8)|^2 \\   + 20|a(1,2,4,6,9)|^2 + 20|a(1,2,4,7,8)|^2 + 2|a(1,3,5,7,9)|^2  }} \nonumber \\
    & = 154.1999084380739947... \,\, .
\end{align}

\end{widetext}

\onecolumngrid
There are six symmetrically non-equivalent non-crossing VB configurations for the ten-site Heisenberg chain ground state, and thus there are six unique coefficients (up to a minus sign) in the VB basis. These six coefficients are given by 
\begin{subequations}
\begin{align}
    \omega & = \frac{4\sqrt{2}}{N_{10}} \Big( |a(1,2,4,7,8)| +3|a(1,2,3,4,5)| - |a(1,2,3,6,7) - 3|a(1,2,3,4,6)| \Big) \nonumber \\
    & = 0.2227271937356899... \,\, , \\
    \nu & = \frac{4\sqrt{2}}{N_{10}} \Big( |a(1,2,3,6,8)| +2|a(1,2,3,4,5)| -2|a(1,2,3,6,7)| \Big) \nonumber \\
    & = 0.0595460383491327... \,\, , \\
    \varepsilon & = \frac{4\sqrt{2}}{N_{10}}|a(1,2,3,4,5)| \nonumber \\
    & = 0.0027237953437734... \,\, ,\\
    \eta & = \frac{4\sqrt{2}}{N_{10}} \Big(|a(1,2,3,4,6)| - |a(1,2,3,4,5)| \Big) \nonumber \\
    & = 0.0136610494273107... \,\, , \\
    \rho & = \frac{4\sqrt{2}}{N_{10}} \Big( |a(1,2,3,6,7) + |a(1,2,3,4,6)| - |a(1,2,3,4,7)| - 2|a(1,2,3,4,5)|   \Big) \nonumber \\
    & = 0.0761178927048571.. \,\, , \\
    \gamma & = \frac{4\sqrt{2}}{N_{10}}\Big(|a(1,2,3,4,7)| + |a(1,2,3,4,5)| - |a(1,2,3,4,6)| \Big) \nonumber \\
    & = 0.0223583966030907... \,\, ,
\end{align}
\end{subequations}
once again in agreement with previous results in the literature \cite{doi:10.1143/JPSJ.58.1403}. Note that the multiplicities of the basis states with coefficients $a(1,2,3,4,7)$ and $a(1,2,3,5,8)$ imply that these BA coefficients are invariant under cyclic permutations, local complement, and one additional symmetry operation which is order inversion. It follows that the ground state of the ten-site chain in the computational basis can be written as 
\begin{align} \label{N10_GS_comp}
    \ket{\text{G.S.}}_{10} & = \frac{1}{N_{10}} \Bigg\{ \sum^{\text{cyc}}_{i \in 1,...,10 } (-1)^i\bigg[ a(1,2,3,4,5)   \ket{\downarrow\downarrow\downarrow\downarrow\downarrow\uparrow\uparrow\uparrow\uparrow\uparrow} + a(1,2,3,4,6)\Big( \ket{\downarrow\downarrow\downarrow\downarrow\uparrow\downarrow\uparrow\uparrow\uparrow\uparrow} - \text{L.C.}\Big) \nonumber \\
    & \kern+0.5em + a(1,2,3,4,7)\Big( \left\{ \ket{\downarrow\downarrow\downarrow\downarrow\uparrow\uparrow\downarrow\uparrow\uparrow\uparrow} - \text{L.C.}\right\} +  \left\{\ket{\uparrow\uparrow\uparrow\downarrow\uparrow\uparrow\downarrow\downarrow\downarrow\downarrow} - \text{L.C.} \right\}\Big) + a(1,2,3,5,7) \Big( \ket{\downarrow\downarrow\downarrow\uparrow\downarrow\uparrow\downarrow\uparrow\uparrow\uparrow} - \text{L.C.}\Big) \nonumber \\
    & \kern+1em + a(1,2,3,5,8) \Big( \left\{\ket{\downarrow\downarrow\downarrow\uparrow\downarrow\uparrow\uparrow\downarrow\uparrow\uparrow} - \text{L.C.} \right\} + \left\{ \ket{\uparrow\uparrow\downarrow\uparrow\uparrow\downarrow\uparrow\downarrow\downarrow\downarrow} - \text{L.C.}\right\}\Big)   + a(1,2,3,5,9)\ket{\downarrow\downarrow\downarrow\uparrow\downarrow\uparrow\uparrow\uparrow\downarrow\uparrow} \nonumber \\
    & \kern+1.5em + a(1,2,3,6,7) \Big(\ket{\downarrow\downarrow\downarrow\uparrow\uparrow\downarrow\downarrow\uparrow\uparrow\uparrow} - \text{L.C.} \Big) + a(1,2,3,6,8)\Big( \ket{\downarrow\downarrow\downarrow\uparrow\uparrow\downarrow\uparrow\downarrow\uparrow\uparrow} - \text{L.C.}\Big)  \nonumber \\
    & \kern+2em + a(1,2,4,5,8)\ket{\downarrow\downarrow\uparrow\downarrow\downarrow\uparrow\uparrow\downarrow\uparrow\uparrow} + a(1,2,4,6,8) \Big( \ket{\downarrow\downarrow\uparrow\downarrow\uparrow\downarrow\uparrow\downarrow\uparrow\uparrow}-\text{L.C.}\Big)  \nonumber \\
    & \kern+2.5em + a(1,2,4,6,9) \Big( \ket{\downarrow\downarrow\uparrow\downarrow\uparrow\downarrow\uparrow\uparrow\downarrow\uparrow} - \text{L.C.} \Big) + a(1,2,4,7,8) \Big( \ket{\downarrow\downarrow\uparrow\downarrow\uparrow\uparrow\downarrow\downarrow\uparrow\uparrow} - \text{L.C.} \Big)\bigg] \nonumber \\
    & \kern+3em + a(1,3,5,7,9)\Big(\ket{\downarrow\uparrow\downarrow\uparrow\downarrow\uparrow\downarrow\uparrow\downarrow\uparrow} - \text{L.C.} \Big) \Bigg\},
\end{align}
\begin{figure*}
    \centering
    \includegraphics[width=\linewidth]{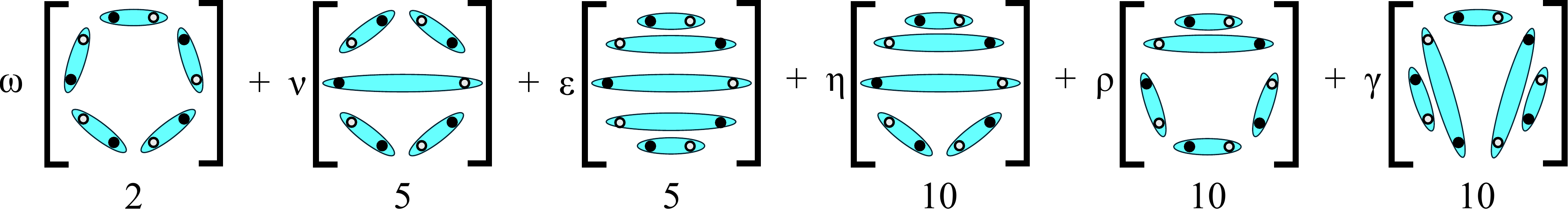}
    \caption{Exact ground state of the ten-site Heisenberg chain in the non-crossing VB basis. Square brackets denote a linear combination over VB configurations and the number below the square brackets is the multiplicity. The coefficients are all Galois unsolvable algebraic numbers and their numeric values are given in the text.}
    \label{fig:N10}
\end{figure*}
and in the VB basis as 
\begin{align} 
    \ket{\text{G.S.}}_{10} & = \omega \Big([1;2][3;4][5;6][7;8][9;10] - [1;10][3;2][5;4][7;6][9;8]\Big) \nonumber \\
    & \kern+2em + \nu \Big( [1;2][3;8][4;5][7;6][9;10] + [1;10][3;2][5;6][7;8][9;4] + [1;2][3;4][5;10][7;6][9;8] \nonumber \\
    & \kern+4em + [1;6][3;2][5;4][7;8][9;10] + [1;10][3;4][5;6][7;2][9;8] \Big) \nonumber \\
    & \kern+2em + \varepsilon \Big( [1;2][3;10][4;9][5;8][7;6] + [1;4][3;2][5;10][7;8][9;6] + [1;6][3;4][5;2][7;10][9;8] \nonumber \\
    & \kern+4em + [1;8][3;6][5;4][7;2][9;10] + [1;10][3;8][5;6][7;4][9;2] \Big) \nonumber \\
    & \kern+2em + \eta \Big( [1;2][3;10][5;6][7;8][9;4] - [1;4][3;2][5;10][7;6][9;8] + [1;6][3;4][5;2][7;8][9;10] \nonumber \\
    & \kern+4em - [1;10][3;6][5;4][7;2][9;8] + [1;2][3;8][5;6][7;4][9;10] - [1;10][3;2][5;8][7;6][9;4] \phantom{\Big(} \nonumber \\
    & \kern+6em + [1;2][3;4][5;10][7;8][9;6] - [1;6][3;2][5;4][7;10][9;8] + [1;8][3;4][5;6][7;2][9;10] \phantom{\Big(} \nonumber \\
    & \kern+8em - [1;10][3;8][5;4][7;6][9;2] \Big)\nonumber \\
    & \kern+2em + \rho \Big( [1;2][3;10][5;4][7;6][9;8] -[1;4][3;2][5;6][7;8][9;10] + [1;10][3;4][5;2][7;6][9;8] \nonumber \\
    & \kern+4em -[1;2][3;6][5;4][7;8][9;10] + [1;10][3;2][5;6][7;4][9;8] - [1;2][3;4][5;8][7;6][9;10] \nonumber \\
    & \kern+6em +[1;10][3;2][5;4][7;8][9;6] - [1;2][3;4][5;6][7;10][9;8] + [1;8][3;2][5;4][7;6][9;10] \nonumber \\
    & \kern+8em -[1;10][3;4][5;6][7;8][9;2] \Big) \nonumber \\
    & \kern+2em + \gamma \Big( [1;2][3;10][5;4][7;8][9;6] -[1;4][3;2][5;6][7;10][9;8] + [1;8][3;4][5;2][7;6][9;10] \nonumber \\
    & \kern+4em - [1;10][3;6][5;4][7;8][9;2] + [1;2][3;10][5;6][7;4][9;8] - [1;4][3;2][5;8][7;6][9;10] \nonumber \\
    & \kern+6em + [1;10][3;4][5;2][7;8][9;6] - [1;2][3;6][5;4][7;10][9;8] + [1;8][3;2][5;6][7;4][9;10] \nonumber \\
    & \kern+8em - [1;10][3;4][5;8][7;6][9;2]\Big). 
\end{align}

\section{$N = 12$ Heisenberg Chain}

Here we compute a high-precision numeric reconstruction of the twelve-site Heisenberg chain ground-state wavefunction. The $N=12$ case is potentially solvable within the coordinate Bethe ansatz in a manner similar to the previous sections; however, the computation is prohibitively expensive. As a result, $N=12$ is the first chain size for which the algebraic Bethe ansatz becomes imperative for deriving the exact ground state. 
\subsection{Algebraic BA}

The $N=12$ ground state Baxter polynomial contains three $Q$-coefficients and is given by 
\begin{equation}
    Q(u) = u^6 + c_4 u^4 + c_2 u^2 + c_0. 
\end{equation}
Baxter polynomiality yields a system of three polynomials in the variables $c_4,c_2,$ and $c_0$. For brevity, we do not include the full remainder, polynomial system, or full univariate polynomial of the Gr\"obner basis eliminations here, but they are available upon reasonable request. Eliminating all but one $Q$-coefficient and factorizing the univariate polynomials of the Gr\"obner basis yields the following polynomials:
\begin{subequations}
    \begin{align}
        P^{12}(c_4) & = 6879707136 c_4^{10}-74530160640 c_4^9+246379511808 c_4^8-264582070272 c_4^7+7469604864 c_4^6 \nonumber \\
        & \kern+1em +116662726656 c_4^5-27515211264 c_4^4-13716288000 c_4^3+1698407632 c_4^2+714269352 c_4+45952497 \\ 
        P^{12}(c_2) & = 22543424343244800000 c_2^{10}-168512096965754880000 c_2^9+320187465254043648000 c_2^8 \nonumber \\
        & \kern+1em -103620367293087744000 c_2^7-155216773359410872320 c_2^6+70753912689719771136 c_2^5 \nonumber \\
        & \kern+2em +262938817641185280 c_2^4 -2147657771275026432 c_2^3+19298522728075520 c_2^2 \nonumber \\
        & \kern+3em +3337015342942368 c_2 -42054587228451\\ 
        P^{12}(c_0) & = 496408368206987447500800000 c_0^{10}+427708995821199006105600000 c_0^9 \nonumber \\
        & \kern+1em -54145462670649955713024000 c_0^8-7427167404969116565504000 c_0^7 \nonumber \\
        & \kern+2em -142356489476505096683520 c_0^6-277026657860755390464 c_0^5+12491070331944960 c_0^4 \nonumber \\
        & \kern+4em +11647844352000 c_0^3-1424338944 c_0^2+32640 c_0+1. 
    \end{align}
\end{subequations}
The physical solutions for the ground state are given by the roots $P_1^{12}(c_4)$, $P_5^{12}(c_2)$, and $P_5^{12}(c_0)$. Substituting these roots into the Baxter polynomial and computing the resultant yields the $N=12$ Bethe rapidity polynomial, which we do not include here due to its cumbersome size. Notably, it is even and of degree sixty, which is six times larger than the $Q$-coefficient degree. The phase cosine polynomial of the $N=12$ ground state, $P^{12}(x)$, is found to be 
\begin{align}
P^{12}(x) & = 1073741824 x^{30}-9663676416 x^{29}+50734301184 x^{28}-180388626432 x^{27}+326618841088 x^{26} \nonumber \\
& \kern+1em +109253230592 x^{25} -1735468777472 x^{24}+2886553567232 x^{23}+400568614912 x^{22}-7658111238144 x^{21} \nonumber \\
& \kern+2em +8846185594880 x^{20} +3983735259136 x^{19}-18148337647616 x^{18}+11000761810944 x^{17} \nonumber \\ 
& \kern+3em +13067031019520 x^{16}-21370456113152 x^{15} +2089686515712 x^{14}+16091279720448 x^{13} \nonumber \\
& \kern+4em -10159076515840 x^{12} -4915457032192 x^{11}+7452365195264 x^{10} -639627203584 x^{9} \nonumber \\
& \kern+5em -2600843335936 x^{8} +943712814080 x^{7}+429499178560 x^6-271791222656 x^5 \nonumber \\
& \kern+6em -17144289376 x^4+31543356288 x^3-2851693564 x^2-1129363332 x+174539673,
\end{align}
where the physical solutions are given by $P_{11}^{12}(x) = \cos{\theta_{16}}$, $P_9^{12}(x) = \cos{\theta_{25}} $, and $P_{6}^{12}(x) = \cos{\theta_{34}} $. Notably, this polynomial has triple the degree of the $Q$-coefficient polynomials. The Galois group of $P^{12}(x)$ is found to be $S_3 \wr S_{10}$ with order $6^{10} \cdot 10! = 219419659468800$. As $S_{10}$ is an unsolvable group the $N=12$ cosine phases are not expressible in radicals. 
\subsection{Coordinate BA}
Here, we show the phase equations required to perform the coordinate Bethe ansatz derivation for the $N=12$ Heisenberg chain ground state. Ultimately, it is not feasible to compute the cosine phase polynomial analytically via this method due to expression swell. The Bethe quantum numbers for the twelve-site chain are $\{1,3,5,7,9,11\}$ and we note that $k_1 = \theta_{16}$, $k_2 = \theta_{25}$, and $k_3 = \theta_{34}$. The ground-state energy is parametrised by the three aforementioned phase angles and three (arbitrarily chosen) auxiliary phase angles, for this example we choose $\{\theta_{13}, \theta_{14}, \theta_{23} \}$. The six phase angle equations thus take the form: 
\begin{subequations}
\begin{align}
0 & = -  \sin(\frac{1}{4} (2 \theta_{14} - 13 \theta_{16} - 2 \theta_{23} + 9 \theta_{25} - 11 \theta_{34})) + 2\sin(\frac{1}{4} (2 \theta_{14} - 9 \theta_{16} - 2 \theta_{23} + 9 \theta_{25} - 11 \theta_{34})) \nonumber \\
& \kern+2em -  \sin(\frac{1}{4} (2 \theta_{14} - 9 \theta_{16} - 2 \theta_{23} + 13 \theta_{25} - 11 \theta_{34})), \label{N12_long_a}\\
0 & = \cos(\frac{1}{2}(\theta_{16} - \theta_{25})) \sin(\frac{1}{4} (2 \theta_{13} - 11 \theta_{16} + 2 \theta_{23} - 11 \theta_{25} + 11 \theta_{34})) - \sin(\frac{1}{4} (2 \theta_{13} - 9 \theta_{16} + 2 \theta_{23} - 9 \theta_{25} + 11 \theta_{34})), \label{N12_long_b} \\
0 & = \cos(\frac{1}{2} (\theta_{13} - \theta_{16} - \theta_{34})) - 
  2 \cos(\frac{1}{2} (\theta_{13} - \theta_{16} + \theta_{34})) + \cos(\frac{1}{2}(\theta_{13} + \theta_{16} + \theta_{34})), \\
0 & = \cos(\frac{\theta_{14}}{2}) \cos(\frac{1}{2}(\theta_{16} - \theta_{34})) - \cos(\frac{1}{2} (\theta_{14} - \theta_{16} - \theta_{34})), \\
0 & = \cos(\frac{1}{2} (\theta_{13} - \theta_{14} + \theta_{23} - \theta_{25} + 10 \theta_{34})) - \cos(\frac{1}{2}(\theta_{25} - \theta_{34})) \cos(\frac{1}{2} (\theta_{13} - \theta_{14} + \theta_{23} + 11 \theta_{34})), \\ 
0 & =  \cos(\frac{1}{2} (\theta_{23} - \theta_{25} - \theta_{34})) - 2 \cos(\frac{1}{2} (\theta_{23} - \theta_{25} + \theta_{34})) + \cos(\frac{1}{2} (\theta_{23} + \theta_{25} + \theta_{34})).
\end{align}
\end{subequations}

\begin{figure*}
    \centering
    \includegraphics[width=0.75\linewidth]{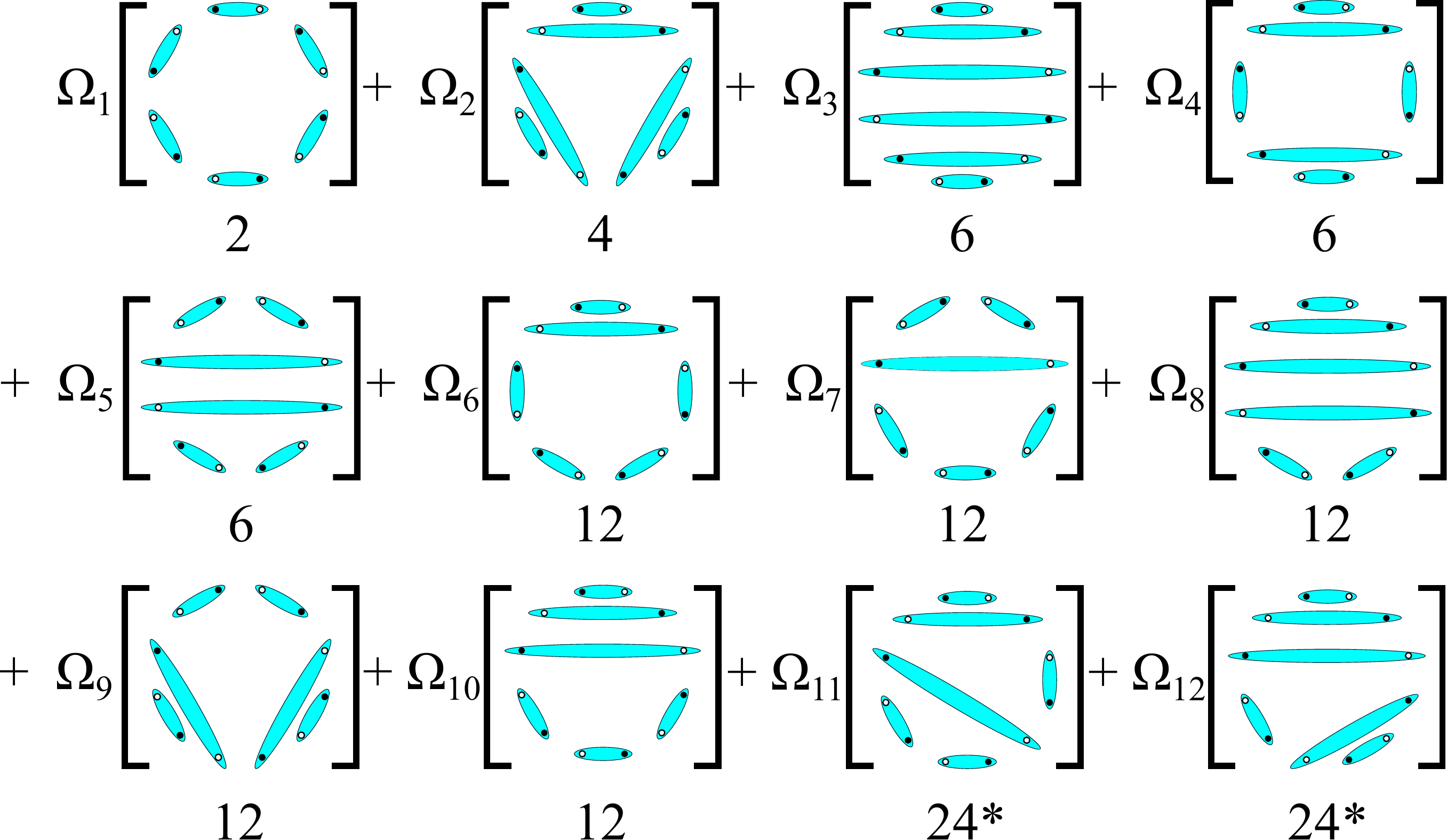}
    \caption{Exact ground state of the twelve-site Heisenberg chain in the non-crossing VB basis. Square brackets denote a linear combination over VB configurations and the number below the square brackets is the multiplicity. The two configurations marked with an asterisk ($*$) are the first to demonstrate chirality, whereby they are rotationally non-equivalent to their reflections yet possess the same coefficients. }
    \label{fig:N12}
\end{figure*}
To continue in the same fashion as the $N=10$ case via the tangent half-angle substitution requires the trigonometric expansion of all functions in terms of a single phase angle each. However, expanding \cref{N12_long_a,N12_long_b} both yield expressions with 9120 terms each and the resulting Gr\"obner basis calculation is computationally intractable. Consequently, $N=12$ becomes the fist size of the Heisenberg chain ground state for which the coordinate Bethe ansatz cannot be performed with comparable computational ease to the algebraic Bethe ansatz.  
\subsection{Ground state quantities}

\begin{table}
\caption{\label{N12_table} Numeric values of the unique ground state Bethe ansatz coefficients of the $N=12$ antiferromagnetic Heisenberg chain and their multiplicities.  }
\begin{tabular}{c|c|c} 
Coefficient & Numeric value & Multiplicity \\
\hline
a(1,2,3,4,5,6) & -0.0115742497495828 & 12\\
a(1,2,3,4,5,7) & 0.0855035074763103 & 24\\
a(1,2,3,4,5,8) & -0.2435495313692128 & 48\\
a(1,2,3,4,5,9) & 0.3392405472849705 & 24\\
a(1,2,3,4,6,8) & 1.1302808782052322 & 24\\
a(1,2,3,4,6,9) & -1.9233124591879486 & 48\\
a(1,2,3,4,6,10) & 1.5562671962649787 & 48\\
a(1,2,3,4,6,11) & -0.5936153416397770 & 12\\
a(1,2,3,4,7,8) & -1.1089567626644792 & 24\\
a(1,2,3,4,7,9) & 3.5528884806802905 & 48\\
a(1,2,3,4,7,10) & -3.5986033590186748 & 24\\
a(1,2,3,4,8,9) & -1.7465911529488527 & 24\\
a(1,2,3,5,6,9) & 3.9564381910619890 & 24\\
a(1,2,3,5,6,10) & -3.6735096943051939 & 24\\
a(1,2,3,5,7,9) & -12.6499916575356362 & 24 \\
a(1,2,3,5,7,10) & 14.2779273749197489 & 48 \\
a(1,2,3,5,7,11) & -6.7358491310309163 & 24 \\
a(1,2,3,5,8,9) & 7.6032719861553616 & 48 \\
a(1,2,3,5,8,10) & -17.2541428716697967 & 48 \\
a(1,2,3,5,8,11) & 10.1881568545792547 & 24 \\
a(1,2,3,5,9,10) & 5.9306228677992528 & 24 \\
a(1,2,3,6,7,10) & -10.7740379542570403 & 48 \\
a(1,2,3,6,8,9) & -9.5416837796734358 & 24 \\
a(1,2,3,6,8,10) & 25.7825609045945110 & 24 \\
a(1,2,3,7,8,9) & 2.9876623813999682 & 6\\
a(1,2,4,5,7,10) & -18.7861223873994777 & 24\\
a(1,2,4,5,8,10) & 27.8896078447174379 & 24\\
a(1,2,4,6,8,10) & -75.5483089356955369 & 24\\
a(1,2,4,6,8,11) & 56.1763031334861606 & 24\\
a(1,2,4,6,9,10) & 34.6039174597622987 & 24\\
a(1,2,4,6,9,11) & -51.8134407141136194 & 12\\
a(1,2,4,7,8,10) & 33.0502329005941085 & 12\\
a(1,2,4,7,8,11) & -29.9571494660731709 & 24\\
a(1,2,5,6,9,10) & -19.2693929158042661 & 4\\
a(1,3,5,7,9,11) & 189.8682994485241787 & 2 \\
\end{tabular}
\end{table}

The ground-state energy of the $N=12$ Heisenberg chain is given by 

\begin{equation}
     E_0 = 2\cos \theta_{16} + 2\cos \theta_{25} + 2\cos \theta_{34} - 3.
\end{equation}
Each of the above cosine phase angles that correspond to a magnon momentum is a degree thirty algebraic number. Computing the minimal polynomial of the ground-state energy yields
\begin{equation} \label{N12_Eng}
    P^{12}(E_0) = E_0^{10}+12 E_0^9+42 E_0^8-E_0^7-256 E_0^6-315 E_0^5+323 E_0^4+644 E_0^3+47 E_0^2-189 E_0-27, 
\end{equation}
and we once again see that the ground-state energy has the same algebraic degree as the $Q$-coefficients. Interestingly, when $\frac{N}{2}$ is even the ground-state energy polynomial appears to always be monic, whereas the leading coefficient when $\frac{N}{2}$ is odd is $2^d$ where $d$ is the degree of the ground-state energy polynomial. Construction of the ground-state wavefunction requires the identification of the auxiliary phase angles, which we choose to be $\{\theta_{13}, \theta_{14}, \theta_{23} \}$. Considering \cref{BA_phase_relation,Bethe_roots_to_momenta} it follows that 
\begin{equation}
    \cos(\theta_{ij}) = \frac{(u_i - u_j)^2-1}{(u_i - u_j)^2+1}.
\end{equation}
Noting that the three Bethe rapidities corresponding to the magnon momenta are given by $P_4^{12}(u)$, $P_6^{12}(u)$, and $P_9^{12}(u)$ (up to a negative sign), one could in theory find the minimal polynomial of the auxiliary phase angles in a similar manner to the previous cases. However, we were unable to obtain the minimal polynomial for the auxiliary phase angles within a reasonable computational time, likely due to expression swell. As such, for the $N=12$ auxiliary phase angles we use high precision numerical solutions. They are found to be 
\begin{subequations}
    \begin{align}
    \cos \theta_{13} & = -0.50616434468358905837... \, ,\\
    \cos \theta_{14} & = -0.28986097088524526052... \, ,\\
    \cos \theta_{23} & = -0.92479010613035028733... \, .
    \end{align}
\end{subequations}

The numerical values of the thirty-five BA coefficients and their multiplicities are given in \cref{N12_table}. The $N=12$ ground state normalization factor, for coefficients denoted $a_i$ with corresponding multiplicity $M_i$, is found to be
\begin{align}
    N_{12} & = \sqrt{\sum_{a_i} M_i \abs{a_i}^2 } \nonumber \\
    & = 685.5475389297346856... \, .
\end{align}

Due to the large number of basis states in both the computational and VB basis, for brevity we forgo writing the ground states explicitly and limit our presentation to only the schematic form in \cref{fig:N12}. The twelve symmetrically non-equivalent VB basis states are given by 
\begin{subequations}
    \begin{align}
        \Omega_1 & = \frac{8}{N_{12}}\Big[-a(1,2,5,6,9,10) + 4a(1,2,3,4,5,7) +4a(1,2,3,4,5,9) - 4a(1,2,3,4,6,8) +2a(1,2,3,4,7,8) \nonumber \\
        & \kern+2em  + a(1,2,3,4,5,6) \Big] \nonumber \\
        & = 0.1659141872978956... \, ,\\
        \Omega_2 & = \frac{8}{N_{12}}\Big[-a(1,2,3,4,7,8) + a(1,2,3,4,5,7) + a(1,2,3,4,5,9) - a(1,2,3,4,6,8) + a(1,2,3,4,5,6)\Big] \nonumber\\
        & = 0.0045726449848562... \, ,\\
        \Omega_3 & = -\frac{8}{N_{12}} a(1,2,3,4,5,6) \nonumber\\
        & = 0.0001350657580089... \, , \\
        \Omega_4 & = -\frac{8}{N_{12}}\Big[\frac{a(1, 2, 3, 7, 8, 9)}{2} + a(1,2,3,4,5,6) \Big] \nonumber\\
        & = 0.0172972038468927... \, , \\
        \Omega_5 & = -\frac{8}{N_{12}}\Big[ a(1, 2, 3, 4, 6, 11) + 2a(1,2,3,4,5,7) + a(1,2,3,4,5,6) \Big]\nonumber\\
        & = 0.0050666954722303... \, , \\
        \Omega_6 & = \frac{8}{N_{12}}\Big[ a(1, 2, 3, 5, 8, 9) + a(1, 2, 3, 4, 6, 9) + a(1, 2, 3, 4, 6, 8) -2a(1,2,3,4,5,7)  -\frac{a(1, 2, 3, 4, 5, 9)}{2} \nonumber \\
        & \kern+2em + a(1, 2, 3, 4, 8, 9) -a(1,2,3,4,5,6) \Big] \nonumber\\
        & = 0.0552503911926498... \, , \\
        \Omega_7 & = \frac{8}{N_{12}}\Big[ \frac{a(1, 2, 3, 5, 8, 11) + a(1, 2, 3, 4, 5, 9)}{2} + a(1, 2, 3, 4, 6, 9)\Big] \nonumber\\
        & = 0.0389806518387812... \, , \\
        \Omega_8 & = -\frac{8}{N_{12}}\Big[ a(1,2,3,4,5,8) + \frac{a(1,2,3,4,5,9)}{2} \Big]\nonumber\\
        & = 0.0008627177959637... \, ,\\
        \Omega_9 & = -\frac{8}{N_{12}}\Big[ a(1, 2, 3, 4, 8, 9) + a(1, 2, 3, 4, 5, 9) \Big]\nonumber\\
        & = 0.0164230840400771... \, , \\
        \Omega_{10} & = \frac{8}{N_{12}}\Big[ -a(1,2,3,4,5,7) - a(1,2,3,4,5,9) + a(1,2,3,4,6,8) \Big]\nonumber\\
        & = 0.0082332650429515... \, , \\
        \Omega_{11} & = \frac{8}{N_{12}}\Big[ -a(1, 2, 3, 4, 6, 9) -a(1, 2, 3, 4, 6, 8) + a(1,2,3,4,5,7) \Big]\nonumber\\
        & = 0.0102520690521983... \, , \\
        \Omega_{12} & = \frac{4}{N_{12}} a(1,2,3,4,5,9) \nonumber\\
        & = 0.0019793845241693... \, . 
    \end{align}
\end{subequations}
The $N=12$ case is the first length where the degree of the minimal polynomial of the ground-state energy is less than the dimension of the VB non-crossing basis subspace up to rotations and reflections. 

\section{Integer sequences and counting basis states}


The Bethe ansatz can be thought of as a powerful symmetry-adapted basis tranformation of the Heisenberg Hamiltonian, thereby block diagonalizing the exponentially large Hilbert space into the smaller, relevant sub-Hilbert spaces that span the energy eigenstates. Particularly within the $S^z$ computational basis, the dimension of the zero-magnetization subspace is $\frac{N!}{[(N/2)!]^2}$, which is substantially smaller than the full Hilbert space of dimension $2^N$. But as we have seen in the previous sections the BA coefficients are invariant up to translations, order inversion, and local complement of the down spin positions. It follows that the number of unique coefficients (up to a minus sign) is given by the number of reversible binary necklace configurations; the enumeration of these configurations is given by the integer sequence A006840 \cite{oeisA006840}. It was this fact that allowed us to write an efficient representation of the exact ground states in the computational basis, as seen in \cref{N6_GS_comp,N8_GS_comp,N10_GS_comp}. However, this is not the most compact form available to the Heisenberg chain ground state. It was first noted by Rumer \cite{rumer1932nachrichten} and later demonstrated by multiple authors \cite{TemperleyLieb1971,doi:10.1143/JPSJ.58.1403,Saito_noncrossing} that any one-dimensional singlet state is completely spanned by the set of non-crossing VB basis states. The dimension of the non-crossing VB basis for the even $N$ site chain is given by the $(N/2)^{\text{th}}$ Catalan number. Within the non-crossing VB basis there are configurations that are equivalent up to rotations and reflections, allowing for a highly compact representation of the ground-state wavefunction. Notably, Hulth\'{e}n used the symmetrically non-equivalent, non-crossing VB basis to approximate the ground-state energy of the chains up to ten sites in 1938 \cite{hulthen1938uber}, long before the advent of the computer or the proof of the completeness of the non-crossing basis. The number of rotationally non-equivalent, non-crossing configurations is enumerated by the number of unlabeled planar trees, given by the integer sequence A002995 \cite{oeisA002995}. The inclusion of reflection symmetry becomes relevant for chains of length twelve and longer, as no VB configurations in the smaller chains are chiral. The number of non-equivalent, non-crossing configurations up to rotations and reflections is given by the number of labelled, projective planar trees, which is integer sequence A006082 \cite{oeisA006082}. The size of the aforementioned subspace dimensions for bipartite Heisenberg chains up to $N=16$ is shown in \cref{Tab_enums}. For the four-site through ten-site cases we have found that the algebraic degree of the ground-state energy matches the number of symmetrically non-equivalent, non-crossing VB states, as first noted by Hulth\'{e}n \cite{hulthen1938uber}. This implies that this basis defines the minimal algebraic structure of the Heisenberg chain ground state; in other words, the Hamiltonian matrix in the non-equivalent, non-crossing VB basis is non-derogatory, by which the minimal polynomial and the characteristic polynomial are equal. We have shown that the first chain length for this pattern to break down is $N=12$; the minimal polynomial of the twelve-site chain is degree ten, thus the characteristic polynomial in this basis factorizes into the degree ten polynomial $P^{12}(E_0)$ in \cref{N12_Eng} and an unknown polynomial of degree two. Thus, it is likely that there exists a subspace of the symmetrically non-equivalent, non-crossing VB basis that characterises the minimal algebraic structure of the Heisenberg chain ground state.

\begin{table}[ht]
\caption{\label{Tab_enums} Dimension of subspaces for the Heisenberg chain ground state up to $N=16$, and the algebraic degree of the ground-state energy. }
\begin{tabular}{c|c|c|c|c|c|c} 
N &$2^N$  & $\genfrac{(}{)}{0pt}{}{N}{N/2}$ \qquad & Catalan Numbers & Binary Necklaces & Labeled Flat Trees & $\deg[P^N(E_0)]$ \\
\hline
    2 & 4 & 2 & 1 & 1 & 1 & 1\\
    4 & 16 & 6 & 2 & 2 & 1 & 1\\
    6 & 64 & 20 & 5 & 3  & 2 & 2\\
    8 & 256 & 70 & 14 & 7  & 3 & 3\\
    10  & 1024 & 252 & 42 & 13  & 6 & 6\\ 
    12  & 4096 & 924 & 132 & 35 & 12 & 10\\ 
    14  & 16384 & 3432 & 429 & 85  & 27 & 20\\ 
    16  & 65536 & 12870 & 1430 & 257  & 65 & 35\\ 
\end{tabular}
\end{table}

\end{document}